\documentclass[aps,onecolumn,preprint,superscriptaddress,nofootinbib,floats]{revtex4}
\usepackage{amsmath,amssymb,color,mathrsfs,graphicx,verbatim,epsf,epsfig,bbm,url,wasysym}
\usepackage[hyperfootnotes=false]{hyperref}
\usepackage{slashed}
\allowdisplaybreaks

\def\beq{\begin{equation}}
\def\eq{\end{equation}}
\def\eeq{\end{equation}}

\def\centeron#1#2{{\setbox0=\hbox{#1}\setbox1=\hbox{#2}\ifdim
\wd1>\wd0\kern.5\wd1\kern-.5\wd0\fi
\copy0\kern-.5\wd0\kern-.5\wd1\copy1\ifdim\wd0>\wd1
\kern.5\wd0\kern-.5\wd1\fi}}
\def\ltap{\;\centeron{\raise.35ex\hbox{$<$}}{\lower.65ex\hbox{$\sim$}}\;}
\def\gtap{\;\centeron{\raise.35ex\hbox{$>$}}{\lower.65ex\hbox{$\sim$}}\;}
\def\gsim{\mathrel{\gtap}}
\def\lsim{\mathrel{\ltap}}

\def\MET{{\not \!  \! E}_T}
\def\vecMET{\vec{\not \!  \! E}_T}

\def\chii0{\chi_i^0}
\def\chij0{\chi_j^0}

\def\gold{\widetilde{G}}

\def\Bino{\widetilde{B}}

\def\higgsino{\tilde{h}^0}
\def\stop{\tilde{t}}
\def\stopR{\tilde{t}_R}
\def\stopL{\tilde{t}_L}
\def\neu{\tilde\chi^0_1}

\def\foursqr#1#2{{\vcenter{\vbox{
 \hrule height.#2pt
 \hbox{\vrule width.#2pt height#1pt \kern#1pt
 \vrule width.#2pt}
 \hrule height.#2pt
 \hrule height.#2pt
 \hbox{\vrule width.#2pt height#1pt \kern#1pt
 \vrule width.#2pt}
 \hrule height.#2pt
     \hrule height.#2pt
 \hbox{\vrule width.#2pt height#1pt \kern#1pt
 \vrule width.#2pt}
 \hrule height.#2pt
     \hrule height.#2pt
 \hbox{\vrule width.#2pt height#1pt \kern#1pt
 \vrule width.#2pt}
 \hrule height.#2pt}}}}
\def\psqr#1#2{{\vcenter{\vbox{\hrule height.#2pt
 \hbox{\vrule width.#2pt height#1pt \kern#1pt
 \vrule width.#2pt}
 \hrule height.#2pt \hrule height.#2pt
 \hbox{\vrule width.#2pt height#1pt \kern#1pt
 \vrule width.#2pt}
 \hrule height.#2pt}}}}
\def\sqr#1#2{{\vcenter{\vbox{\hrule height.#2pt
 \hbox{\vrule width.#2pt height#1pt \kern#1pt
 \vrule width.#2pt}
 \hrule height.#2pt}}}}

\def\figin{\epsfcheck\figin}\def\figins{\epsfcheck\figins}
\def\epsfcheck{\ifx\epsfbox\UnDeFiNeD
\message{(NO epsf.tex, FIGURES WILL BE IGNORED)}
\gdef\figin##1{\vskip2in}\gdef\figins##1{\hskip.5in}
\else\message{(FIGURES WILL BE INCLUDED)}%
\gdef\figin##1{##1}\gdef\figins##1{##1}\fi}
\def\DefWarn#1{}
\def\figinsert{\goodbreak\midinsert}
\def\ifig#1#2#3{\DefWarn#1\xdef#1{fig.~\the\figno}
\writedef{#1\leftbracket fig.\noexpand~\the\figno}%
\figinsert\figin{\centerline{#3}}\medskip\centerline{\vbox{\baselineskip12pt
\advance\hsize by -1truein\noindent\footnotefont{\bf
Fig.~\the\figno:\ } \it#2}}
\bigskip\endinsert\global\advance\figno by1}

\def\fig#1#2#3#4{\vskip 0.5cm \begingroup \midinsert \centerline{
\psfig{file=#1,width=#2}} \vskip 0.4cm
\global\advance\figno by 1
\centerline{\vbox{\baselineskip=12pt \noindent Figure \the\figno: #3}}
\endinsert \endgroup {\xdef#4{\the\figno}} }

\def\figcrop#1#2#3#4#5#6#7#8{\vskip 0.5cm \begingroup \midinsert \centerline{
\psfig{file=#1,width=#2,bbllx=#3,bblly=#4,bburx=#5,bbury=#6}} \vskip 0.4cm
\global\advance\figno by 1
\centerline{\vbox{\baselineskip=12pt \noindent Figure \the\figno: #7}}
\endinsert \endgroup {\xdef#8{\the\figno}} \vskip .5cm}

\def\figlabel#1{\xdef#1{\the\figno}}
\def\encadremath#1{\vbox{\hrule\hbox{\vrule\kern8pt\vbox{\kern8pt
\hbox{$\displaystyle #1$}\kern8pt}
\kern8pt\vrule}\hrule}}
\def\underarrow#1{\vbox{\ialign{##\crcr$\hfil\displaystyle
 {#1}\hfil$\crcr\noalign{\kern1pt\nointerlineskip}$\longrightarrow$\crcr}}}

\begin{document}

\begin{titlepage}
\thispagestyle{empty}

\begin{center}
\vspace*{-1cm}

\hfill UTTG-20-12 \\
\hfill TCC-019-12 \\
\vskip 0.65in
{\LARGE \bf Cornering Light Stops with Dileptonic $\mathbf{m_{T2}}$} \\
\vspace{.15in}

\vskip 0.35in
{\large Can Kilic$^1$ and Brock Tweedie$^2$}

\vskip 0.25in
$^1${\em Theory Group, Department of Physics and Texas Cosmology Center \\
The University of Texas at Austin \\
Austin, TX 78712}

\vskip 0.12in
$^2${\em Physics Department \\
Boston University \\
Boston, MA 02215\\}

\vskip 0.4in

\end{center}

\baselineskip=16pt

\noindent  Supersymmetric spectra with a stop NLSP and a neutralino or gravitino LSP present a special challenge for collider searches.  For stop pairs directly produced from QCD, the visible final-state particles are identical to those of top quark pair production, giving very similar kinematics but often with much smaller rates.  The situation is exacerbated for compressed spectra with $m_{\stop} \simeq m_t + m_{\rm LSP}$, as well as for lighter stops which can suffer from low acceptance efficiencies.  In this note, we explore the power of a direct stop search using dileptonic $m_{T2}$, similar to the one recently performed by ATLAS, but more optimized to cover these difficult regions of the $(m_{\stop},m_{\rm LSP})$ plane.  Our study accounts for the effects of stop chirality and LSP identity, which can be significant.  In particular, our estimates suggest that $m_{\stop} \simeq m_t$ with a massless LSP is excludable for right-handed stops with bino-like (gravitino) LSP with 2012 (2011) data, but remains largely unobservable in the case of a higgsino-like singlino LSP.  For each of these cases we map out the regions of parameter space that can be excluded with 2012 data, as well as currently allowed regions that would yield discovery-level significance.  We also comment on the prospects of a precision $m_{T2}$ shape measurement, and consider the potential of ATLAS's dileptonic $\stop \to b\tilde\chi^+_1$ searches when re-interpreted for light stops decaying directly to the LSP.

\end{titlepage}

\baselineskip=17pt

\newpage

\setcounter{page}{1}




\section{Introduction}

Weak-scale supersymmetry (SUSY) is perhaps the most well-studied extension of the Standard Model (SM), since it offers a weakly-coupled solution the to Higgs hierarchy problem, as well as a dark matter candidate and an indication of grand unification at high energies. However, LHC searches have rapidly been excluding ``generic'' superparticle spectra with masses below roughly 1~TeV~\cite{CMSsusy,:2012rz}.  This places significant tension on supersymmetric models, which require in particular that stops should be lighter than about a TeV to avoid the re-introduction of fine-tuning.  In order to preserve the main virtue of supersymmetry while avoiding these constraints, we are led to seriously consider the possibility that third generation squarks are the only colored superparticles in the sub-TeV range~\cite{Cohen:1996vb,Sundrum:2009gv,Brust:2011tb,Papucci:2011wy,Csaki:2012fh,Gherghetta:2011wc,Larsen:2012rq,Craig:2012di}.

Standard SUSY searches rely heavily on the assumption of copious production of colored superparticles and the release of a substantial amount of missing transverse energy ($\MET$) in their cascade decays, ending in a stable and invisible lightest supersymmetric particle (LSP).  In scenarios where the LHC can only access squarks of the third generation, these searches become significantly less sensitive.  To a large extent this is due to the reduction in the number of colored SUSY production channels, and the fact that third generation squarks cannot be efficiently produced through gluino exchange.  While the cross sections become much higher if these squarks are somewhat light, the visible and invisible energy released then tends to be much smaller, and the signal can easily be missed in searches tailored to more generic spectra.  This has motivated a new interest in dedicated searches for light third generation squarks.

One of the simplest and best-motivated channels for such a search is QCD pair production of the lightest stop eigenstate, with each stop decaying into a top quark and the LSP.  The particles in the final state are then identical to those in top quark pair production, supplemented by the two invisible LSP's.  If there is a large mass gap between the stop and the LSP, stop production events can be distinguished from SM $t\bar t$ production by identifying kinematically extreme events:  events with large $\MET$ in the all-hadronic channel, events with large transverse mass $m_T(l,\MET)$ in $l$+jets channel, and events with large ``stransverse mass'' $m_{T2}(l^+,l^-,\MET)$~\cite{Lester:1999tx,Barr:2003rg,Burns:2008va} in the dileptonic channel.  Indeed, several LHC searches are already capitalizing on these strategies, and are beginning to set meaningful limits~\cite{:2012si,:2012ar,Aad:2012uu,Chatrchyan:2012wa,CMSrazor,CMSbtagRazor,CMSmultijetRazor}.  Nonetheless, there remain regions of parameter space that are not well-covered by these searches, either because the missing energy release is too small (the {\it compressed} case $m_{\stop} \simeq m_t + m_{\rm LSP}$) or because the visible energy release is too small (the {\it off-shell} case $m_{\stop} < m_t + m_{\rm LSP}$).  For example, in the case of $m_{\stop} \simeq m_t$ and a nearly-massless LSP, the final-state kinematics become very similar to SM $t\bar t$ production, but with cross sections that are roughly six times smaller and which rapidly decrease as the stop and LSP masses are raised together.  Lighter stops can have much larger cross sections, but also much smaller acceptances, and the latter can outweigh the former depending on exactly how a search is defined.

In this paper, we will study the kinematics of events arising from compressed spectra and spectra with off-shell decays, and survey regions of the $(m_{\stop},m_{\rm LSP})$ plane where signal sensitivity could be enhanced with a more optimized search strategy using dileptonic $m_{T2}$.  Optimization of light stop searches is a task that has recently been taken up by several other groups, which used a diverse set of approaches.  In~\cite{Alves:2012ft}, a data-driven, shape-based analysis of the $\MET$ and $m_T$ distributions of all-hadronic and $l$+jets channels was considered.  Good sensitivity in the compressed limit was claimed for 2012, with exclusion up to $m_{\stop} \simeq 325$~GeV.  However, this study effectively considered statistical errors only, and not the possible systematic uncertainties involved in extrapolating from control-region fits to signal-region predictions.  Nor is it clear how to define control regions that are enriched in tops but depleted in stops at high-$\MET$ or high-$m_T$, as would be necessary to obtain a precise background-only shape prediction from the data.  Ref.~\cite{Han:2012fw} also considered a high-precision kinematic discrimination, but instead of exploiting the different decay kinematics of stops versus tops, differences in production kinematics and spin correlations were used.  The former was revealed using the rapidity distributions of the top/stop decay products, and the latter was revealed using a matrix-element style discriminator.  In the dileptonic channel, these methods were shown to yield 2--3$\sigma$ discrimination between SM top production and tops plus stops by the end of 2012, for the specific case of $m_{\stop} \simeq 200$~GeV and $m_{\neu} \simeq 0$.  These results were also obtained in the absence of experimental systematics, but they are expected to be quite stable against theory systematics.  The case of stops decaying to a gravitino LSP in gauge mediation was considered in~\cite{Chou:1999zb,Kats:2011it,Kats:2011qh}.  Using the results of various Tevatron and LHC searches (up to 1~fb$^{-1}$ at 7~TeV) not dedicated to this particular process, exclusion potential for $m_{\stop} \lsim m_{t}$ was claimed.  These studies also discussed the potential of dedicated searches using observables such as $m(l,b)$.  Additional studies that consider strategies to improve stop searches, but do not specifically address compressed or off-shell cases, can be found in~\cite{Kaplan:2012gd,Bai:2012gs,Plehn:2012pr}.

To this long list of preceding work, we seek to add the following.  First, we will design a simple ``cut and count'' style search with the dileptonic $m_{T2}$ variable, optimized for compressed spectra, that allows for the presence of nonvanishing systematic errors.  For the recent ATLAS searches these errors tend to be at the $O$(10\%) level, and we will assume systematics of this size.  As we will show, our strategy also automatically yields sensitivity to off-shell decays.  Second, we will investigate the impact of the model-dependence that is present even in this highly simplified SUSY spectrum.  ATLAS searches for stops decaying directly to the LSP always assume a right-handed chiral stop and a bino-like neutralino.  Most theory papers make the same assumption, or in some cases assume a gravitino LSP.  For our search, we will also consider what happens if left-handed stops contribute, or if the stop decays into a light Higgsino-like neutralino without a nearby chargino, as can easily arise if a light singlino mixes into the Higgs sector.  We will show that we can achieve very good exclusion coverage in some cases, while others are more difficult.  We will also discuss how these model-dependent effects become relevant for searches in the $l$+jets channel.

While ATLAS has already performed a search using dileptonic $m_{T2}$~\cite{Aad:2012uu}, that search was optimized for finding stops well above the compressed regime, where a large amount of $\MET$ is released.  Thus an $m_{T2}$ cut significantly beyond $m_W$ could still have reasonable signal acceptance.  Currently, the lowest excluded $m_{\stop}$ with vanishing neutralino mass from this search is 280~GeV.  We propose simple modifications which greatly enhance the potential to exclude or discover stops at lower masses, even those in the compressed limit and below. We will argue that applying these changes to the ATLAS 2011 dileptonic analysis can extend their exclusion boundary at $m_{\neu} = 0$ from 280~GeV down to 210~GeV, and simultaneously exclude most of the region $m_{\stop} \lsim m_t$ that is still allowed by LEP.  This still leaves a gap in exclusion-level sensitivity between 175 and 210~GeV, but we will show that this gap can be completely closed using 2012 data.  The gap inevitably re-opens when we consider more massive neutralinos.  We estimate that its width is typically 5$\sim$10~GeV, and will shrink further when the 14~TeV LHC comes online.  Our study also reveals regions in the $(m_{\stop},m_{\neu})$ plane with discovery-level potential near and below the compressed limit, which are currently not excluded.  As a supplemental analysis, we re-interpret ATLAS's light $\stop \to b\tilde\chi^+_1$ searches, which can cover direct $\stop \to bl\nu\neu$ in complementary regions where the stop/LSP mass splitting is too small to produce an on-shell $W$.

These results apply to the nominal ATLAS model with a right-handed stop decaying into a bino-like neutralino.  When we consider more general models, the $m_{T2}$ distributions can change significantly due to a combination of momentum-scaling and spin effects in the decay matrix elements.  We find that the most promising case is actually a right-handed stop decaying to gravitino, $\stopR \to t\gold$, for which we achieve excellent coverage for essentially arbitrary stop masses below 380~GeV using only 2011 data.  If we instead look at $\stopL \to t\gold$, then the $m_{T2}$ analysis becomes less effective and a gap remains between $m_{\tilde{t}}=190$~GeV and 240~GeV even with the full 2012 dataset.  The most difficult cases are a right-handed stop decaying to a higgsino-like LSP ($\stopR \to t\higgsino$), or a left-handed stop decaying to a bino-like LSP ($\stopL \to t\Bino$).  These are phenomenologically equivalent, and have the softest $m_{T2}$ spectra amongst our models, in some cases even softer than $t\bar t$ itself.  For these models, stops of less than 240~GeV with a massless LSP cannot be excluded with the 2012 data using our method.  Compressed spectra must then be covered using other precision measurements, such as those proposed by~\cite{Alves:2012ft,Han:2012fw}.

The good results obtainable with a simple cut-and-count style search suggest that dileptonic $m_{T2}$ would itself be an excellent candidate for a precision shape-based measurement.  Neglecting systematics and again focusing on a right-handed stop decaying to bino-like neutralino, we estimate that stops of any mass near the compressed limit could be ruled out for $m_{\neu} \lsim 40$~GeV by the end of 2012, and up to nearly 200~GeV after the energy upgrade.

The outline of the paper is as follows.  In section~\ref{sec:kinematics}, we review the kinematic variable $m_{T2}$ and motivate its use in our search.  In section~\ref{sec:decay}, we discuss the physics of stop decay in more detail.  We show how nontrivial distortions of $m_{T2}$ distributions persist even in the compressed limit, and explain their model-dependence.  In section~\ref{sec:analysis}, we begin by reproducing the results of the ATLAS dileptonic $m_{T2}$ analysis in order to validate our methods, and proceed to show how a modified set of cuts can greatly expand the reach for light stops.  In section~\ref{sec:results}, we evaluate the prospects of our search strategy for all choices of stop chirality and LSP identity, and consider some of its limitations and possible extensions.  We summarize our findings and conclude in section \ref{sec:conclusions}.  Two appendices describe in more detail our simulations and statistical procedures.

\section{Dileptonic $m_{T2}$ in Stop Searches}
\label{sec:kinematics}

For each of the three main $t\bar t$ decay channels, we can construct a simple kinematic observable that, under ideal circumstances, exhibits an edge or endpoint.  In the case of all-hadronic decay, this observable is simply $\MET$, and the ``edge'' is at zero.  For $l$+jets decay, the observable is the transverse mass of the lepton-$\MET$ system, $m_T(l,\MET)$.  This is the mass obtained by projecting the entire event into the transverse plane, pretending that the $\vecMET$ vector arises from a single massless particle, and taking the lepton-$\MET$ system mass in 2+1 dimensions.  In events with a single leptonic $W$, $m_T$ is automatically smaller than the full $W$ mass, and has a endpoint at $m_W$.  This endpoint is sculpted into a sharply-peaked edge by a phase space Jacobian factor.

For the dileptonic decay, the analog of $m_T$ is the ``stransverse mass'' $m_{T2}(l^+,l^-,\MET)$~\cite{Lester:1999tx,Barr:2003rg,Burns:2008va}, built out of the subsystem consisting of the two leptons and $\MET$.  To motivate the construction of $m_{T2}$, we assume that the measurable $(l^+,l^-,\MET)$ system arises from two identical one-step decay chains $X^+ \to l^+\nu$ and $X^- \to l^-\bar\nu$.  If it were somehow possible to find the correct decomposition of $\vecMET$ into into its individual neutrino contributions, and break the combinatoric ambiguity over which neutrino to assign to which lepton, we could estimate the mass of the hypothetical parent $X$ by again constructing $m_T$'s for both sides of the event.  Of course the correct $\vecMET$ decomposition is not available in practice, however there is a procedure to obtain a {\it lower bound} for the individual $m_T$'s on an event-by-event basis. This is achieved by considering all possible partitions  $\vec{p}_T(\nu)+\vec{p}_T(\bar\nu) = \vecMET$, and computing the following extremum:
\beq
m_{T2}(l^+,l^-,\MET) \; \equiv \;  \underset{ \vec{p}_T(\nu)+\vec{p}_T(\bar\nu)  \,=\;\, \vec{\not \! E}_T}{\min}  \left[ \max \{m_T(l^+,\nu),m_T(l^-,\bar\nu)\} \right].
\label{eq:mt2opt}
\eeq
In events which genuinely contain two on-shell $W$'s serving as our $X$'s, and no additional sources of $\MET$, the correct transverse $\nu$ and $\bar\nu$ configuration yields $m_T(l^+,\nu)$ and $m_T(l^-,\bar\nu)$ that are both smaller than $m_W$.  Since the above procedure by definition gives a number that is less than or equal to the larger of these two true $m_T$'s, $m_{T2}$ is also strictly bounded above by $m_W$.  The endpoint at $m_W$ is not technically an edge, since the distribution falls continuously to zero over a finite range, but the approach is nonetheless quite steep.

When all final state particles are massless, there is an analytic procedure to compute $m_{T2}$~\cite{Lester:2011nj}.  The trajectories of the leptons in the transverse plane can be viewed as defining the boundary of two wedges, one of which has an opening angle less than $\pi$.  If the $\MET$ vector lies within this wedge, then the configuration that extremizes equation (\ref{eq:mt2opt}) is the one where the two neutrinos move collinearly to their sister leptons, and in this configuration $m_{T2}$ is identically zero. Otherwise, the extremization problem can be reduced to finding the roots of a quartic equation.  The extremized configuration has $\Delta\phi(\nu,\bar\nu) = -\Delta\phi(l^+,l^-)$ and $m_T(l^+,\nu) = m_T(l^-,\bar\nu)$.  In the special case where $\vecMET$ exactly balances the $\vec{p}_{T}$ of the two leptons, such as we would get with two $W$'s with equal and opposite $\vec{p}_T$'s (including two $W$'s at rest), the solution is just $(\vec{p}_T(\nu),\vec{p}_T(\bar\nu)) = (-\vec{p}_T(l^-),-\vec{p}_T(l^+))$.  Thus in this special kinematic configuration:
\beq
m_{T2} \;\to\; \sqrt{2p_T(l^+)\,p_T(l^-)\,(1 + \cos\Delta\phi(l^+,l^-))}.
\eeq
The largest $m_{T2}$ values, near $m_W$, are obtained when both $W$ decays are fully transverse to the beampipe, and the two leptons are collinear and balanced against the two neutrinos.  While derived for a special case, this still gives an approximate picture of the kinematics as long as the $W^+W^-$ system is not itself highly boosted in the transverse plane.

A stop search can utilize any of the three final states and the corresponding observables mentioned above:  $\MET$ for all-hadronic, $m_T$ for $l$+jets, and $m_{T2}$ for dileptonic. The two LSP's produced in the stop decays inject additional $\MET$ into the event, allowing values beyond the edges/endpoints in each of these observables to be populated, in principle leading to significant enhancements in signal over background.

Of course, taking the realities of a hadron collider environment into consideration, the purification of $S/B$ in these three channels is not equally practical.  All-hadronic $t\bar t$ events have $\MET$ from sources such as jet energy mismeasurement and $B$-meson decays, and due to contamination from $t\bar t$ decaying into $l$+jets or even dileptonic final states (especially those with taus). It is therefore quite common to observe all-hadronic events with 10's of GeV of $\MET$, which cause a significant background for genuine signal events, especially in compressed spectra.  Nonetheless, all-hadronic searches are possibly the best option for heavier stops~\cite{:2012si,Chatrchyan:2012wa,CMSrazor,CMSbtagRazor,CMSmultijetRazor}, and it will be important to fully understand how far these searches might be pushed into the compressed regime.

The next-best strategy is to conduct a search in the $l$+jets channel, looking for events with $m_T > m_W$.  However, a search in this channel also suffers from $t\bar t$ backgrounds beyond the kinematic edge, mainly due to misidentified dileptonic events with a hadronic tau. The situation is nonetheless much better than for all-hadronic, and ATLAS exclusion limits based on the semileptonic final state currently extend down to $m_{\stop} \simeq 225$~GeV for a massless bino-like LSP and right-handed stop~\cite{:2012ar}.  But this result is already systematics limited, and a significant improvement will likely require better modeling of the dileptonic contamination.  Another possibility, already utilized in the all-hadronic search, is to place an explicit tau veto, perhaps supplemented with a soft lepton veto.\footnote{We estimate that approximately 2/3 of ATLAS's high-$m_T$ events are dileptonic $t\bar t$ with a tau.  Most of the rest are prompt dileptonic events with a lost lepton.}  We do not explore that option here, as it is difficult for us to reliably model.

If we are simply interested in a basic cut-and-count analysis with reasonable $S/B$, that leaves us with the dileptonic channel and the $m_{T2}$ variable. Of the three different variables and decay channels, this one has the most well-defined endpoint with the fewest outliers, as has clearly been observed experimentally~\cite{Aad:2012uu}.  Events with one or more taus are harmless, as only leptonic taus can contribute, and these tend to populate lower $m_{T2}$ values.  The few background events with $m_{T2}$ far beyond $m_W$ tend to arise either from $t\bar t$ events with mismeasured $\MET$, or from SM electroweak backgrounds and $t\bar t W/Z$.

\begin{figure}
   \includegraphics[width=3in]{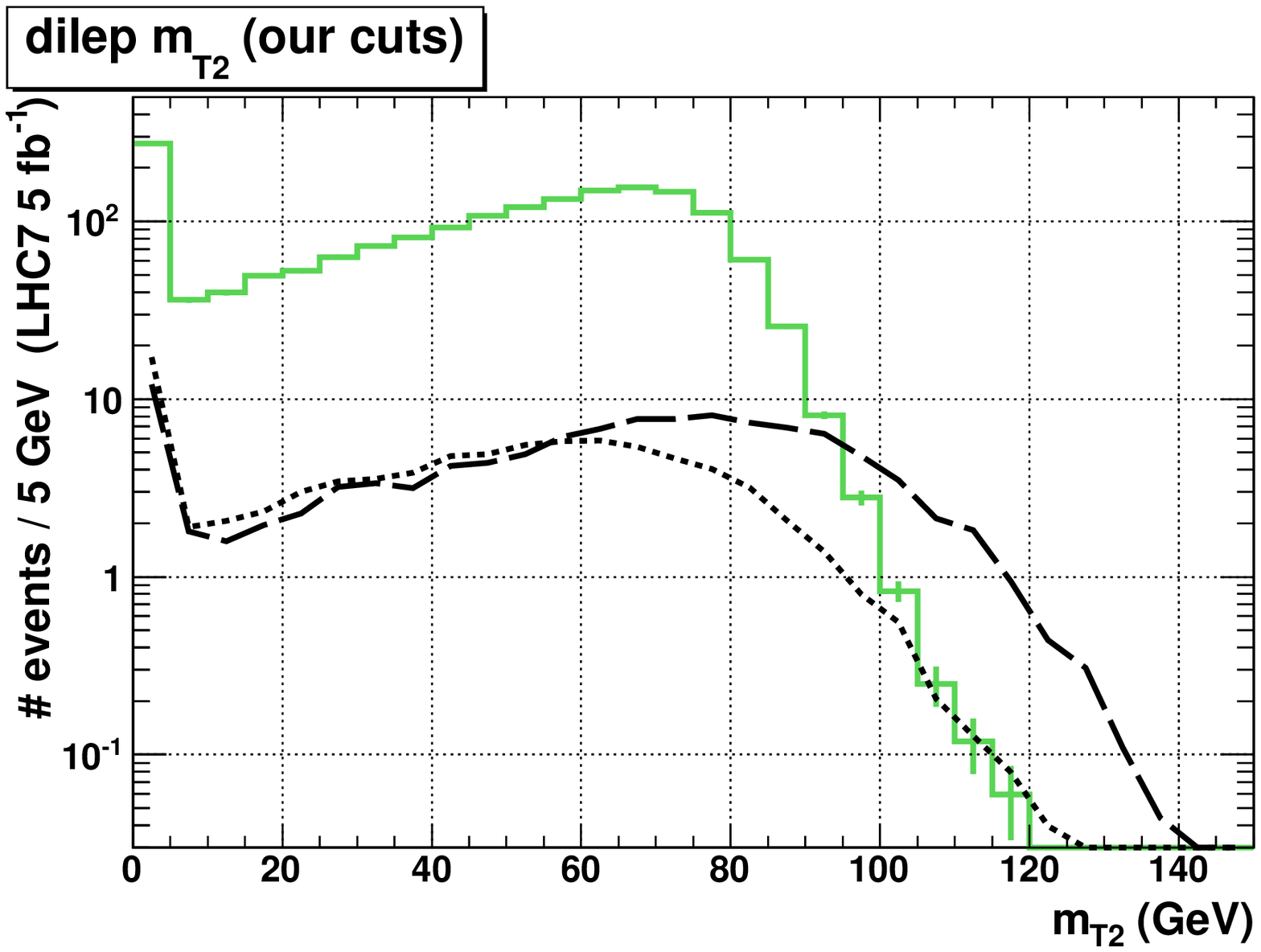}
   \includegraphics[width=3in]{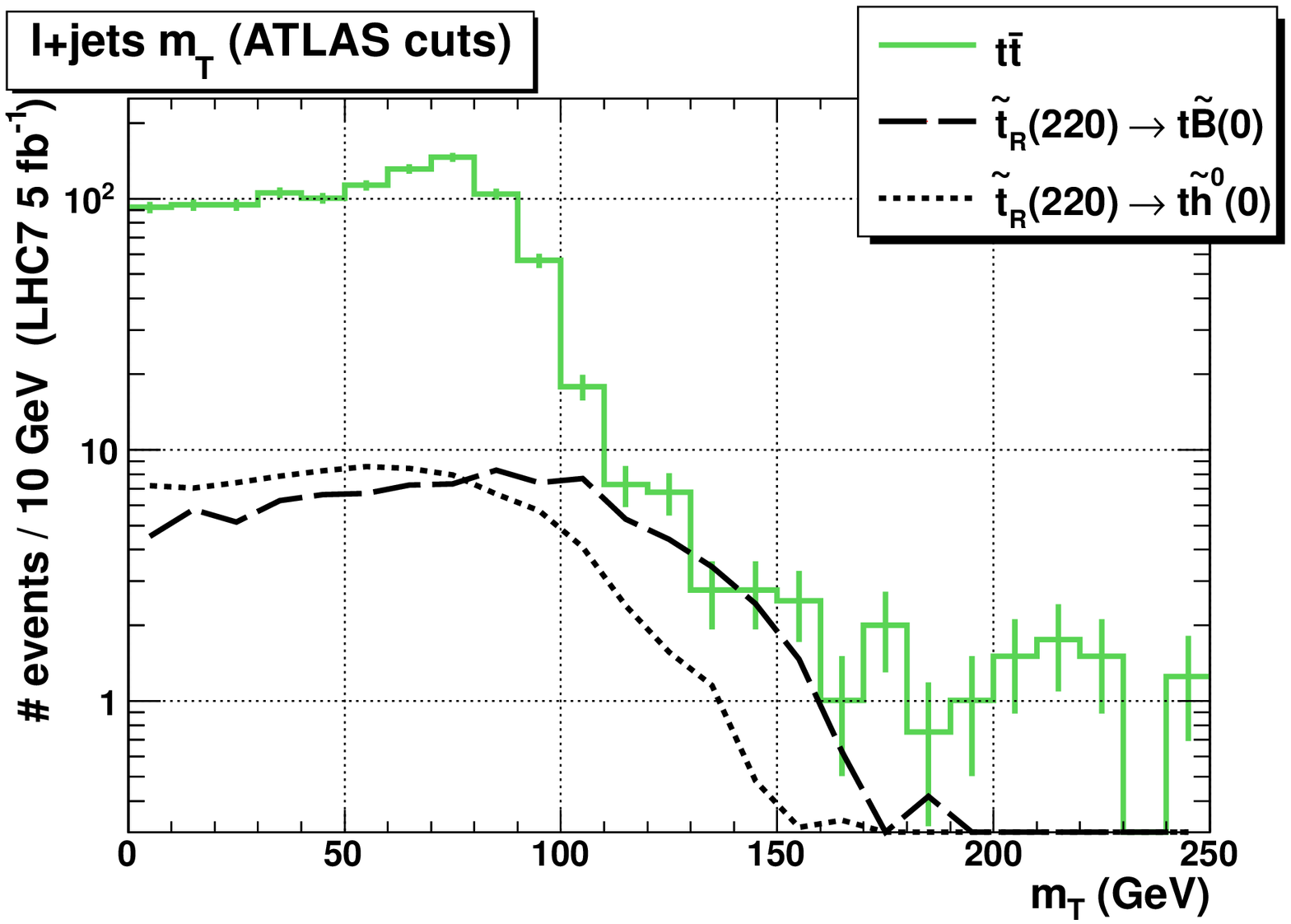}
\caption{The distribution of dileptonic $m_{T2}$ (left) and semileptonic $m_{T}$ (right) for 220~GeV right-handed stops decaying into either a massless bino-like neutralino or a massless higgsino-like neutralino.  For comparison, the distribution for the SM $t\bar t$ background is also shown.  Events have been processed through the reconstruction described in appendix~\ref{sec:simulation}.  The $m_{T2}$ distribution was formed after application of our final analysis cuts described in section~\ref{sec:oursearch}.  The $m_T$ distribution was formed after application of the ``SR~A'' type cuts described in~\cite{:2012ar}, except for the cut on $m_T$ itself.  (Error bars are Monte Carlo statistics.)}
\label{fig:left-vs-right}
\end{figure}

In Fig.~\ref{fig:left-vs-right}, we show fully reconstructed dileptonic $m_{T2}$ and semileptonic $m_T$ distributions for two very moderately compressed example models (just below the current ATLAS $l$+jets exclusion) and for the dominant background $t\bar t$.  The $m_{T2}$ distribution uses cuts that we describe in section~\ref{sec:oursearch}, while the $m_T$ distribution uses cuts defined by ATLAS for their lowest-mass search region~\cite{:2012ar}.\footnote{We have verified that our own $l$+jets analysis faithfully reproduces that of ATLAS.  In particular, we obtain the same shape of $m_T$ with preselection cuts, and accurately predict the $t\bar t$ counts in all five of their signal regions.}  The former clearly has a significantly better $S/B$.  The difference between the models is due to spin effects, which we will shortly discuss in detail.

\section{Stop Decay}
\label{sec:decay}

\subsection{Basic kinematics}

In a simplified SUSY spectrum containing only the lightest stop eigenstate and an invisible LSP, stops decay into the same final states as top quarks, but with the addition of the LSP.  In high-scale mediation models with a neutralino LSP, a semileptonic stop decay looks like $\stop \to bl^+\nu\neu$.  In low-scale mediation models with a gravitino LSP, the analog is $\stop \to bl^+\nu\gold$.  If the stop is heavy enough relative to the LSP, namely $m_{\stop} > m_t + m_{\rm LSP}$, it can simply undergo a 2-body decay $\stop \to t(\neu/\gold)$ followed by the decay of the on-shell top quark.  Indeed, to date, two body decays are always assumed in LHC searches for stop pairs.  However, if the stop is somewhat lighter, but still heavy enough to produce an on-shell $W$ boson ($m_{W}+m_b+m_{\rm LSP} < m_{\stop} < m_t + m_{\rm LSP}$), it will instead undergo a 3-body decay $\stop \to bW^+(\neu/\gold)$.  Even lighter stops, with $m_{\stop} < m_{W}+m_b+m_{\rm LSP}$, will naively undergo a fully 4-body decay.\footnote{Often, it is assumed that the loop- and flavor-suppressed 2-body decay $\stop \to c\neu$ takes over in this region in the case of neutralino LSP.  But the relative branching fraction between this mode and the flavor-conserving 4-body decay is actually highly model-dependent, and it is not difficult to make the flavor-violating mode subdominant~\cite{Boehm:1999tr}.  It is also possible for the stop to undergo a 3-body or 4-body decay through an off-shell chargino or an off-shell sfermion if either of these is not much heavier than the stop. We will not explicitly consider any of these contributions in this work, though we expect that the off-shell chargino/sfermion contributions would not significantly alter our conclusions if present.}  Ultimately, as the available phase space is completely closed off, the stop becomes quasi-stable, a case which we do not consider.

As has been pointed out before~\cite{Kats:2011it,Han:2012fw}, the transitions between the different $N$-body decay regions are not perfectly abrupt.  For example, consider a maximally compressed spectrum with a massless LSP and a stop infinitesimally heavier than the top quark.  The phase space for the 2-body decay is nearly vanishing, naively providing us with an LSP with zero 4-momentum and a final state that looks exactly like a top quark.  However, in reality the 3-body decay (with $m(W,b) < m_t$) will have much more phase space available in this regime, and will dominate.  The LSP energy then is no longer constrained by the small splitting $m_{\stop}-m_t$, but by the much larger splitting $m_{\stop}-m_{W}-m_b$.  If we increase the stop mass, the phase space for the 2-body decay begins to open up, and the distribution of LSP energies starts to become localized near $m_{\stop}-m_t$.   But the details of this transition depend sensitively on the structure of the decay amplitudes.  For $\neu$ LSP, the squared amplitudes are directly proportional to the $\neu$ energy.  This acts to suppress the 2-body decay relative to the 3-body, since the latter is still capable of producing $\neu$ with much more energy.  In fact, the 3-body decay can contribute $\gsim \!\!\! 10\%$ of the branching fraction up to $O$(10~GeV) above the threshold, at least for a massless or very light $\neu$.  For the $\gold$ LSP, which is always practically massless, the decay amplitude carries an additional momentum factor due to the derivative coupling of a goldstino.  As a consequence, the suppression of soft LSP production is much stronger, and the transition region is much broader.

For somewhat more massive neutralinos, this crossover from 3-body to 2-body decays occurs more rapidly as a function of $m_{\stop}$.  Partially this is because the spatial momentum available to the 2-body decay starts to open up much faster than in the nearly-massless case (the scaling switches from $p \sim (m_{\stop}-m_t-m_{\neu})$ to $p \sim \sqrt{m_{\neu}(m_{\stop}-m_t-m_{\neu})}$), but also because the $\neu$ energy factor that appears in the matrix elements becomes bounded from below by $m_{\neu}$.  However, this narrower transition to 2-body decays, where the neutralinos take up a smaller fraction of the available energy, is partially offset by the fact that a small kinetic energy carried by a massive particle can still correspond to a sizeable momentum (a point strongly emphasized in~\cite{Alves:2012ft}).  Even right at the threshold, the approximately $1.3$~GeV intrinsic width of the top quark can lead to a nontrivial momentum dispersion, for example at the $O$(10~GeV) level if the neutralino mass is $O$(100~GeV).  The resulting contributions to the missing momentum can therefore remain observably large even for very compressed spectra.

The fact that there is really no way to completely squeeze out the LSP's momentum, even as we pass through the 2-body/3-body threshold, is advantageous for our $m_{T2}$-based search as well as for all other searches that target the LSP's effects on the decay kinematics.  Far above the threshold $m_{\stop} = m_t + m_{\rm LSP}$, the two LSP's in a stop pair production event carry a large amount of invisible momentum and allow $m_{T2}$ to sometimes exceed $m_W$.  At and below this threshold, the decays are 3-body with an on-shell $W$ and a large fraction of the energy going to the LSP, again allowing for $m_{T2} > m_W$.  Somewhere slightly above the threshold, the $\MET$ reaches a minimum, and extreme values of $m_{T2}$ become much less common.  This region is the most challenging one, but we will show that it is still often possible to obtain good sensitivity there.

If we consider even lower stop masses, approaching the 4-body threshold, the rate of off-shell $W$'s rapidly climbs.  Despite the added $\MET$, the $m_{T2}$ distribution inevitably becomes softer and the endpoint eventually falls below $m_W$.  This presents an absolute end to the utility of the $m_{T2}$ search (as well as the $m_T$ search in the $l$+jets channel).  Nonetheless, other search strategies are already being employed that cover part of this region, and we will describe these in subsequent sections.

Before proceeding, we also comment on the stop decay lifetime.  The decays are often rapid enough to be considered prompt, but not always.  The simplest counterexample is gauge mediation with a moderately high $F_{\rm SUSY}$.  For example, taking $\sqrt{F_{\rm SUSY}} \gsim 100$~TeV, a 200~GeV stop can propagate more than $O$(mm) distance before decaying~\cite{Chou:1999zb,Kats:2011it}.  For stops lighter than 130~GeV, decay lengths are at least on the order of a mm unless $\sqrt{F_{\rm SUSY}} \lsim 10$~TeV.  Stop decays to a neutralino LSP would usually proceed much faster, unless $\neu$ is dominantly singlino, in which case the lifetime can be made arbitrarily long by adjusting the singlino-gaugino-higgsino mixing.  In this paper, we will only consider prompt decays and focus on traditional search strategies.  This assumption can be satisfied in the bulk of the parameter space, except for the region $m_{\stop} \lsim 100$~GeV for a gravitino LSP, which in any case should have been highly visible at LEP.

The region $m_{\stop} + m_{\rm LSP} < m_t$ involves an additional constraint from top quarks decaying into $\stop(\neu/\gold)$.  However, there is a broad range of values for the stop-top-LSP couplings such that the branching fraction of this top decay mode is very small, while the stop decay remains prompt.  This happens automatically in gauge mediation, and a pure bino LSP would be marginally safe.  A pure higgsino-like LSP would probably be ruled out, but even a modest mixing with a singlino state is sufficient to avoid the constraint in this case as well.

\subsection{Spin effects}  \label{sec:spin}

We have seen above that the momentum spectrum of the LSP in stop decay can depend on whether it is a neutralino or a gravitino, and (in the case of $\neu$) on its mass.  When we construct quantities like $m_{T2}(l^+,l^-,\MET)$ or $m_T(l,\MET)$ in a complete event, we must also pay attention to the kinematics of the leptons and neutrinos, and their correlations with the LSP's.  This is especially true since we are looking for events that lie on the tails of distributions, and, as we now explain, these tails are heavily affected by the spin of the top along the stop decay axis.\footnote{ For a more detailed discussion of spin effects in stop decay, as well as their possible measurement at the LHC, see~\cite{Perelstein:2008zt}.}

Let us first consider the case of a gravitino LSP.  The couplings of the gravitino preserve chirality, so, for example, right-handed chirality stops couple to right-handed chirality top quarks.  We are considering decays far below the limit where the tops are produced relativistically, so seemingly the top chirality and helicity are not simply related.  However, the gravitino {\it is} relativistic, and the chirality structure of the interaction vertex completely governs its helicity.  Right-handed chirality stops therefore always produce right-handed helicity gravitinos.  Since the stop is spin-zero, the top quark is also forced to have right-handed helicity, regardless of its velocity.  In other words, the top quarks are 100\% polarized along the decay axis in right-handed chiral stop decay.  In fact, this occurs even if the top is off-shell.  For left-handed chiral stops, the same argument goes through, up to a subdominant contribution from the $\stop_Lb_LW\gold$ 4-point supercurrent coupling.

The fact that the tops produced in chiral stop decays are polarized leads to a significant effect on the final state kinematics.  The angular distribution of the lepton produced in a top quark decay is maximally correlated with the top's spin.\footnote{$d\Gamma/d\Omega(l) \propto 1 + \langle \hat S(t) \rangle\cdot\hat\Omega(l)$, where $\hat\Omega(l)$ is the lepton's unit vector and $\langle \hat S(t) \rangle$ is the top's normalized spin vector, both in the top's rest frame.}  Therefore, in the parent stop's rest frame, the lepton tends to follow the direction of the top in right-handed decays, and tends to follow the direction of the gravitino in left-handed decays.  There is also a similar bias for the neutrino, but it is less pronounced and goes in the opposite direction.  The net effect in right-handed stop decays is that the missing energy contributions from the neutrino and gravitino tend to reinforce each other, and the entire invisible subsystem tends to be back-to-back with the lepton.  This is an optimal situation for enhancing $m_{T2}$ in the dileptonic final state and $m_T$ in the $l$+jets final state in stop pair events.  In contrast, the opposite tendency for left-handed stops tends to {\it suppress} $m_{T2}$ and $m_T$.  For our own $m_{T2}$ search, we therefore expect that searches for right-handed stops decaying to gravitinos will be more sensitive than for left-handed stops.

Similar effects occur with a neutralino LSP.  However, there we have more choices, and the behavior is somewhat more complicated.  First, neutralino couplings to stops come in two varieties:  chirality-preserving gaugino couplings and chirality-flipping higgsino couplings.  Restricting ourselves momentarily to $m_{\neu} = 0$, we expect exactly the same spin effects as in the gravitino LSP case, but with a reversed sense of chirality for a neutralino that couples dominantly like a higgsino.\footnote{There is a slight subtlety if the neutralino couples like a very pure wino.  Then even a mostly-$\stop_R$ will be forced to decay via its $\stop_L$ component.  Though the coupling then technically preserves chirality, it will appear to flip it.  In any case, our very general treatment covers even this extreme situation.}  There is therefore a correspondence between switching the LSP identity $(\neu \sim \Bino) \leftrightarrow (\neu \sim \higgsino)$ and switching the stop chirality $\stop_R \leftrightarrow \stop_L$.  When the neutralino is massive, but still couples dominantly like a gaugino or higgsino, the strength of the polarization depends on its velocity in the decay:  fast-moving neutralinos are again approximately chiral and force the top to be polarized, while slow-moving neutralinos and their associated tops are unpolarized.  The latter occurs exactly in the limit of compressed spectra, though the depolarization is always tempered by the presence of a nontrivial 3-body contribution.

Given these observations, we can identify four distinct limiting cases for the kinematics of stop decay. These are:
\begin{itemize}
\item $\stop_{\rm R} \to t\gold$
\item $\stop_{\rm L} \to t\gold$
\item $\stop_{\rm R} \to t\Bino$ \, or \, $\stop_{\rm L} \to t\higgsino$
\item $\stop_{\rm L} \to t\Bino$ \, or \, $\stop_{\rm R} \to t\higgsino$
\end{itemize}
The different $m_{T2}$ distributions arising from the last two cases, for the specific example of 220~GeV stops decaying to massless neutralinos, is illustrated in Fig.~\ref{fig:left-vs-right}.

To fully understand the possible impact of the spin and momentum structure of the stop decay amplitudes, we will study the LHC sensitivity of each limiting case independently.  While we expect that this captures most of the relevant phenomenology, a given spectrum can of course entail more complicated patterns of neutralino and stop mixings, and interference effects could become quite nontrivial.  There are nonetheless some further simplifications that can occur.  For example, if the LSP is massless and pure gaugino, higgsino, or gravitino, but the stop is arbitrarily mixed, the contributions from the $\stop_R$ and $\stop_L$ components do not interfere.  Also, if the LSP is couples like a mostly-bino, it preferentially selects out the $\stop_R$ component due to the larger hypercharge.

\section{Search Strategy}
\label{sec:analysis}

In this section we will describe stop search strategies based on the use of dileptonic $m_{T2}$.  We begin by describing the ATLAS search on which we base our own, and reproducing the kinematic distributions therein.  This analysis was optimized for spectra far from the compressed regime and was therefore sensitive to heavier stops.  Nevertheless, understanding it in detail gives us a chance to validate our Monte Carlo event samples, including important effects such as detector energy resolution.  We then proceed to introduce our own version of this search, which we expect to yield better sensitivity in the compressed regime.

\subsection{The ATLAS dileptonic $m_{T2}$ search}

The recent ATLAS dileptonic $m_{T2}$ search~\cite{Aad:2012uu} is already placing limits on direct production of right-handed stop pairs decaying as $\stop_R \to t\Bino$.  The search uses the following event selection criteria in order to isolate the stop signal in the dileptonic channel:
\begin{itemize}
\item  Exactly two opposite-sign isolated leptons ($l = e$ or $\mu$) satisfying $p_T(e) > 25$~GeV, $p_T(\mu) > 20$~GeV, $|\eta(l)| \lsim 2.5$, and $m(l^+,l^-) > 20$~GeV (regardless of flavors).
\item  At least two jets with $p_T(j) > 25$~GeV and $|\eta(j)| < 2.5$, with the leading jet satisfying $p_T(j_1) > 50$~GeV.
\item  In same-flavor (SF) events, a $Z$-veto ($m(l^+,l^-) \neq [71,111]$~GeV) and at least one $b$-tagged jet.
\item  For the final signal region selection, events with $m_{T2} > 120$~GeV.
\end{itemize}

In order to reproduce the kinematic distributions presented in the ATLAS analysis, as well as for further use in our own analysis, we generate Monte Carlo samples of the signal and the backgrounds for SM $t\bar{t}$, $l^+l^-$+jets, diboson, $t\bar{t}W/Z$ and single top production.  We do not attempt to model fake lepton backgrounds, which are unlikely to be important for us.  The details of the Monte Carlo tools that we use, as well as our approximations of detector effects, are described in detail in appendix~\ref{sec:simulation}.  For the following, we will combine the same-flavor and different-flavor channels into a single analysis, unlike the ATLAS search which keeps them separate.

\begin{figure}
\centering
\begin{center}
   \includegraphics[width=3.15in]{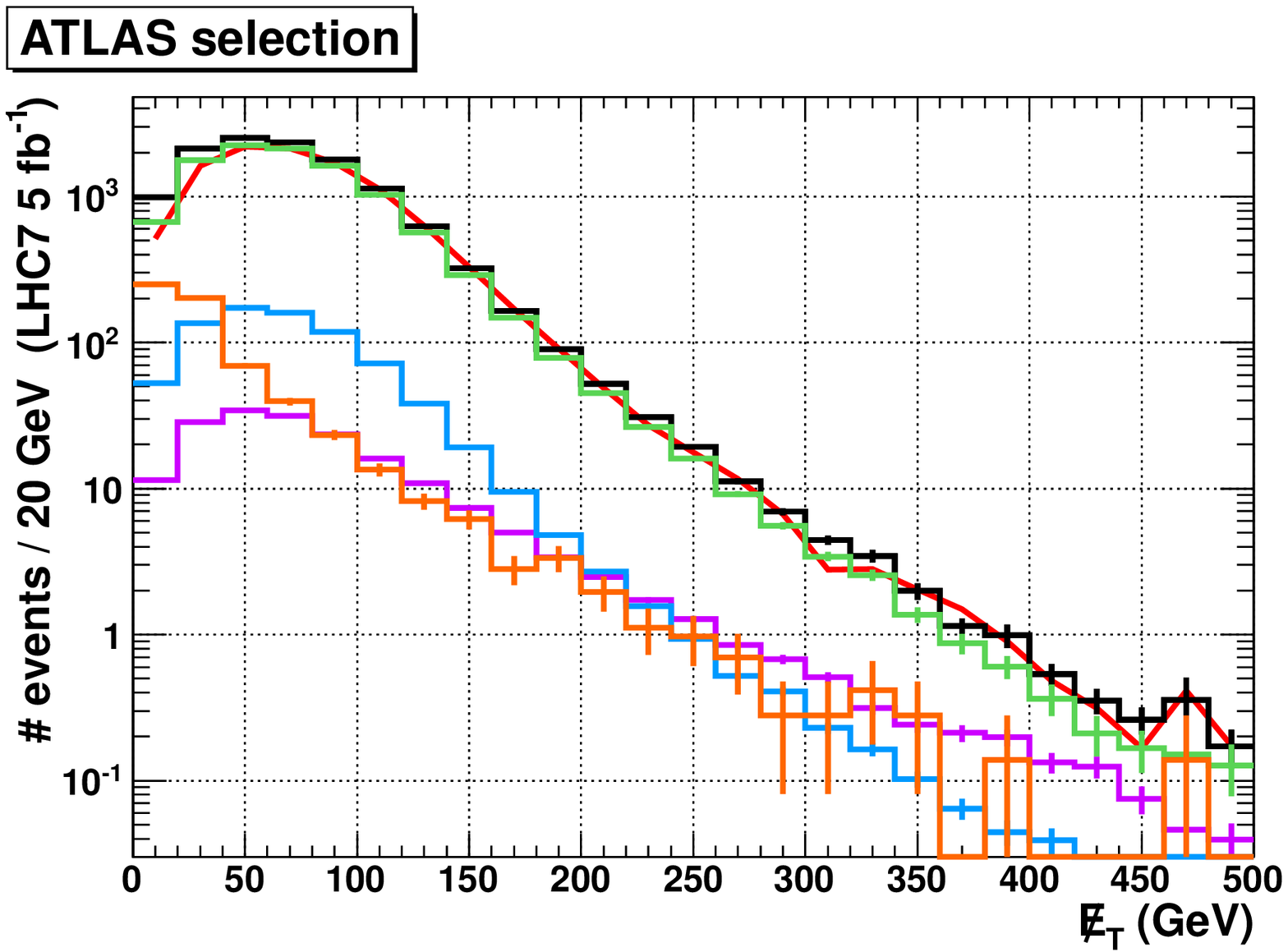}
   \includegraphics[width=3.15in]{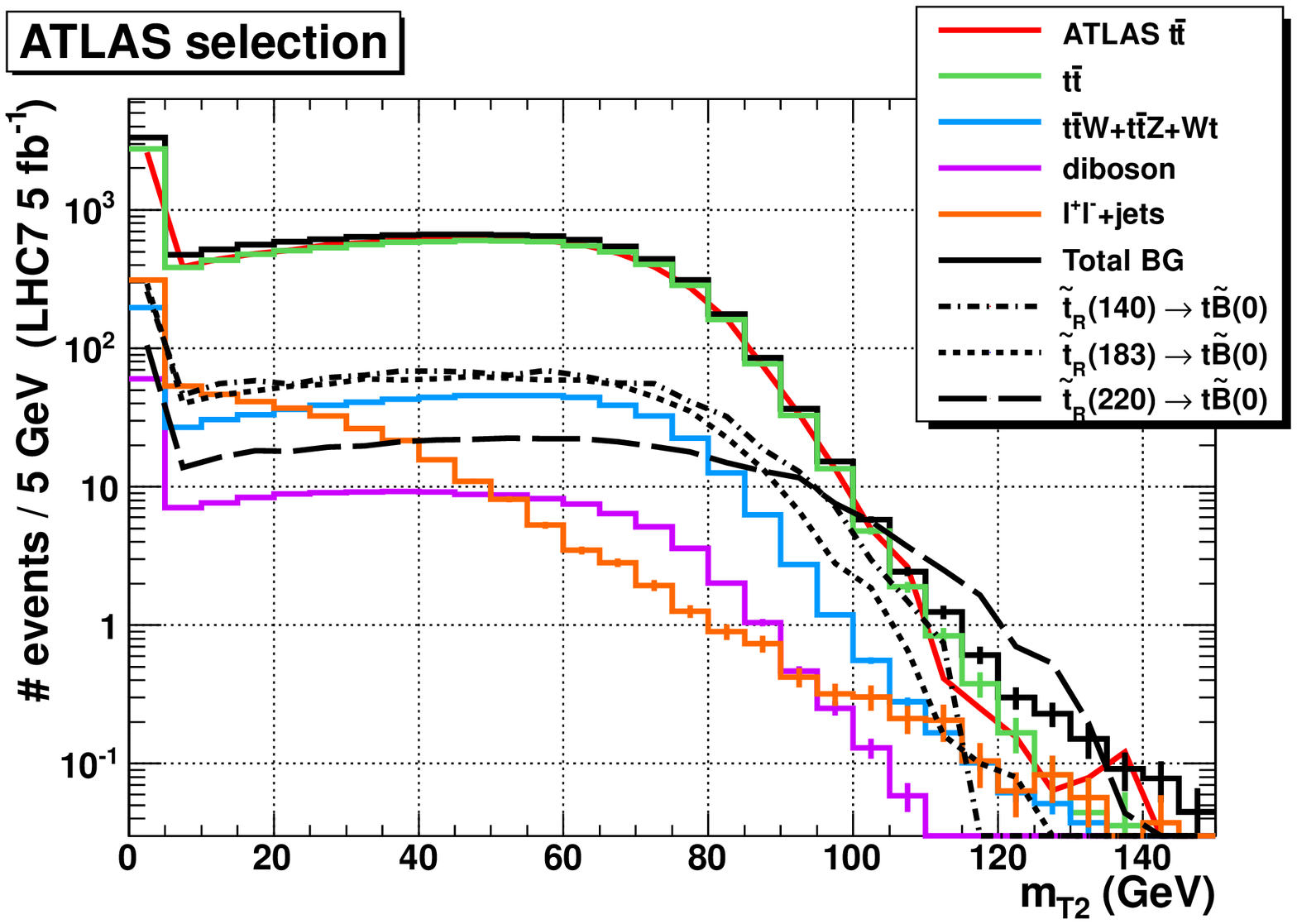}
\end{center}
\caption{$\MET$ (left) and $m_{T2}$ (right) distributions after the selection cuts of the ATLAS analysis have been applied, with 5~fb$^{-1}$ at the 7~TeV LHC.  We show ATLAS's $t\bar t$ prediction (red curve) and our own background predictions:  $t\bar t$ (green), single-top and $t\bar tW/Z$ (blue), electroweak diboson (purple), $l^+l^-$+jets (orange), and total background (black solid). For $m_{T2}$ we also show three signal points, corresponding to a massless bino-like neutralino LSP and right-handed stops with masses of 140~GeV (dot-dashed curve), 183~GeV (dotted curve), and 220~GeV (dashed curve).  (Error bars are Monte Carlo statistics.)}
\label{fig:ATLAScomparison}
\end{figure}

Applying the selection cuts above (except for $m_{T2}$), we display our distributions for $\MET$ and $m_{T2}$ in Fig.~\ref{fig:ATLAScomparison}.  We include the $t\bar{t}$ distribution estimated by the ATLAS analysis for comparison.  In the latter plot, we also show a small set of signal $m_{T2}$ distributions, corresponding to $\stop_R \to t\Bino$ with $m_{\neu} = 0$ and $m_{\stop} = 140$, 183, and 220~GeV.\footnote{Though not shown, we also reproduce the shape and normalization of their example signal point, $(m_{\stop},m_{\neu}) = (300,50)$~GeV.}  The agreement with the ATLAS $t\bar t$ distributions is generally very good, with the largest statistically-significant discrepancies ($O(20\%)$) occurring for the lowest two $\MET$ bins.  These do not contribute to the signal region at large $m_{T2}$.  The largest resolvable discrepancy in modeling of the other backgrounds, which is not a very significant one, is in the $l^+l^-$+jets background.  For $m_{T2} > 120$~GeV we predict roughly $0.5$ event, whereas ATLAS predicts $1.2 \pm 0.5$.  In fact, we will soon show the $l^+l^-$+jets background to be highly subdominant for our own version of the $m_{T2}$ search.

The expected turnover near $m_W$ in the $t\bar t$ background is indeed present, with the endpoint smeared out by finite resolution effects and, to a lesser extent, by the $W$'s natural width.  The falloff is nonetheless rapid, with a roughly two orders of magnitude drop between 60~GeV and 100~GeV, and two more orders of magnitude up to 130~GeV.  Until the turnover point, $t\bar t$ is the leading background, but other backgrounds that do not have as sharp a falloff begin to give an order-one relative contribution at very high $m_{T2}$.  The biggest amongst these are $t\bar t W/Z$ and $l^+l^-$+jets.  The latter can achieve large values of $m_{T2}$ when $O$(10's) of GeV $\MET$ mismeasurement, often supplemented by neutrinos in $b$-hadron decay, points opposite to a highly-boosted $l^+l^-$ system that is off of the $Z$ resonance.

One can see based on Fig.~\ref{fig:ATLAScomparison} why the ATLAS search is not very sensitive to light stops. The signal in the 220~GeV stop mass case exhibits a healthy $S/B$ at high $m_{T2}$, but the expected event count is only $O(2)$ with 2011 LHC data in the region $m_{T2} > 120$~GeV.  This feature should nonetheless become observable with increased statistics.  The 140~GeV stop is much more difficult to observe with an $m_{T2}$ cut alone, although it does achieve $S/B \simeq 1/2$ for a cut near 100~GeV.  The most difficult point, 183~GeV, never exceeds $S/B \simeq 1/4$.


\subsection{Proposed analysis for compressed spectra}
\label{sec:oursearch}

Having validated our analysis setup by reproducing the ATLAS kinematic distributions, we attempt to optimize the dileptonic $m_{T2}$ analysis for the case of light stops with masses in the compressed regime and below.  The most obvious consequence of working in this region of parameter space is a reduction in the available $\MET$ in signal events, and hence a softening of the $m_{T2}$ distribution.  As we saw above, searches for lighter stops can benefit greatly from lowering the final $m_{T2}$ cut, and throughout the rest of our study we will utilize a final cut of 95~GeV.  This particular cut is not always absolutely optimal for all cases, but it is nonetheless quite effective over a broad range of models.  In particular, we have trained our analysis on compressed $\stop_R \to t\Bino$ assuming 25~fb$^{-1}$ of 8~TeV data, and expect this cut to be close to ideal for that case.

In addition to reduced $\MET$ and $m_{T2}$, we must also contend with the softer $b$-jets arising in off-shell decays.  The off-shell decays are a significant contribution for compressed spectra, and can yield some of the highest-$m_{T2}$ events.  Given this fact, it is not obviously beneficial to demand two hard jets in the event selection, especially since one the most efficient ways for the SM to produce hard dileptons with $m_{T2} \gsim m_W$ at the LHC is already $t\bar t$.  Consequently, we largely eliminate any requirements on jets accompanying the two leptons.\footnote{We might also consider using lower $p_T$ cuts on the leptons themselves, and this is in fact possible thanks to the availability of relatively low-threshold dilepton triggers.  However, by demanding high-$m_{T2}$ we are effectively demanding hard leptons anyway, and we do not expect a significant benefit from accepting softer leptons into the analysis.}  A necessary exception, which helps to combat the $ZZ^*$ background, is that we continue to demand at least one $b$-tagged jet in the SF subsample.  Our {\it baseline} selection, before any $m_{T2}$ cut, is therefore almost the same as that of ATLAS, but with the ``At least two jets...'' requirement dropped.

\begin{figure}
\begin{center}
   \includegraphics[width=3.15in]{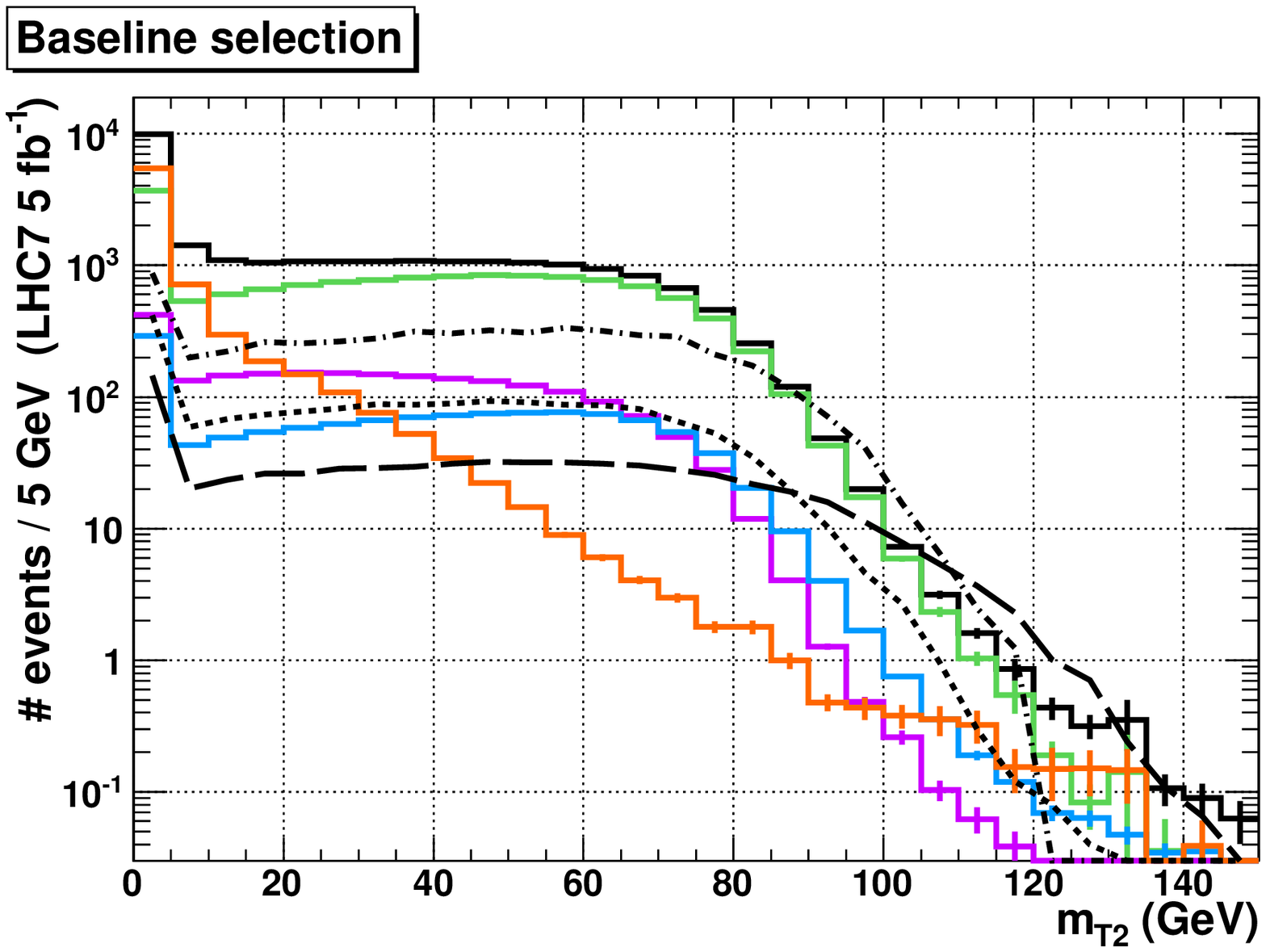}
   \includegraphics[width=3.15in]{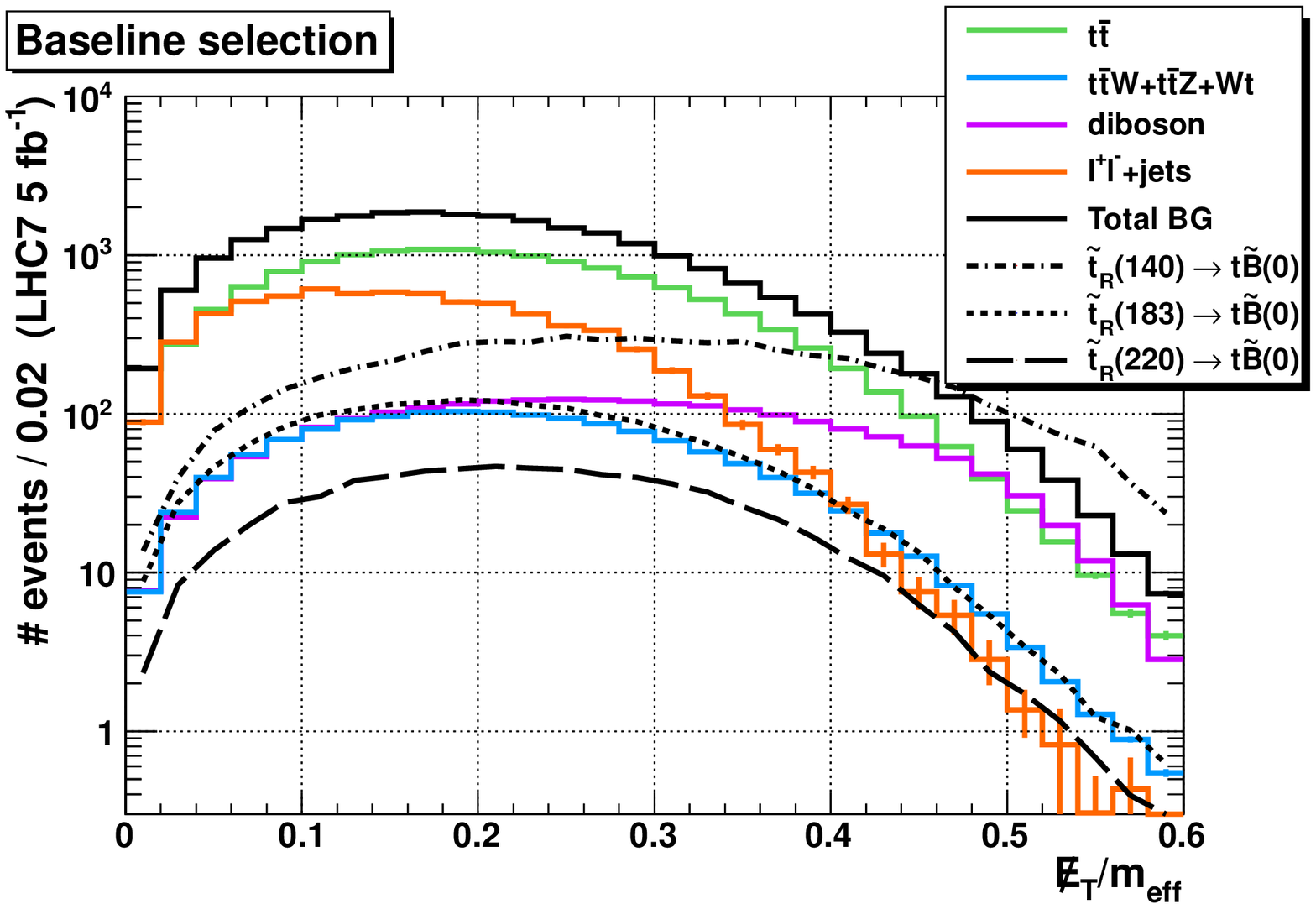}
\end{center}
\caption{$m_{T2}$ (left) and $\MET/m_{\rm eff}$ (right) distributions after our baseline selection cuts, with 5~fb$^{-1}$ at the 7~TeV LHC.  We show our background predictions:  $t\bar t$ (green), single-top and $t\bar tW/Z$ (blue), electroweak diboson (purple), $l^+l^-$+jets (orange), and total background (black solid).  We also show three signal points, corresponding to a massless bino-like neutralino LSP and right-handed stops with masses of 140~GeV (dot-dashed curve), 183~GeV (dotted curve), and 220~GeV (dashed curve).  (Error bars are Monte Carlo statistics.)}
\label{fig:baseline-dist}
\end{figure}

\begin{figure}
\begin{center}
   \includegraphics[width=3.15in]{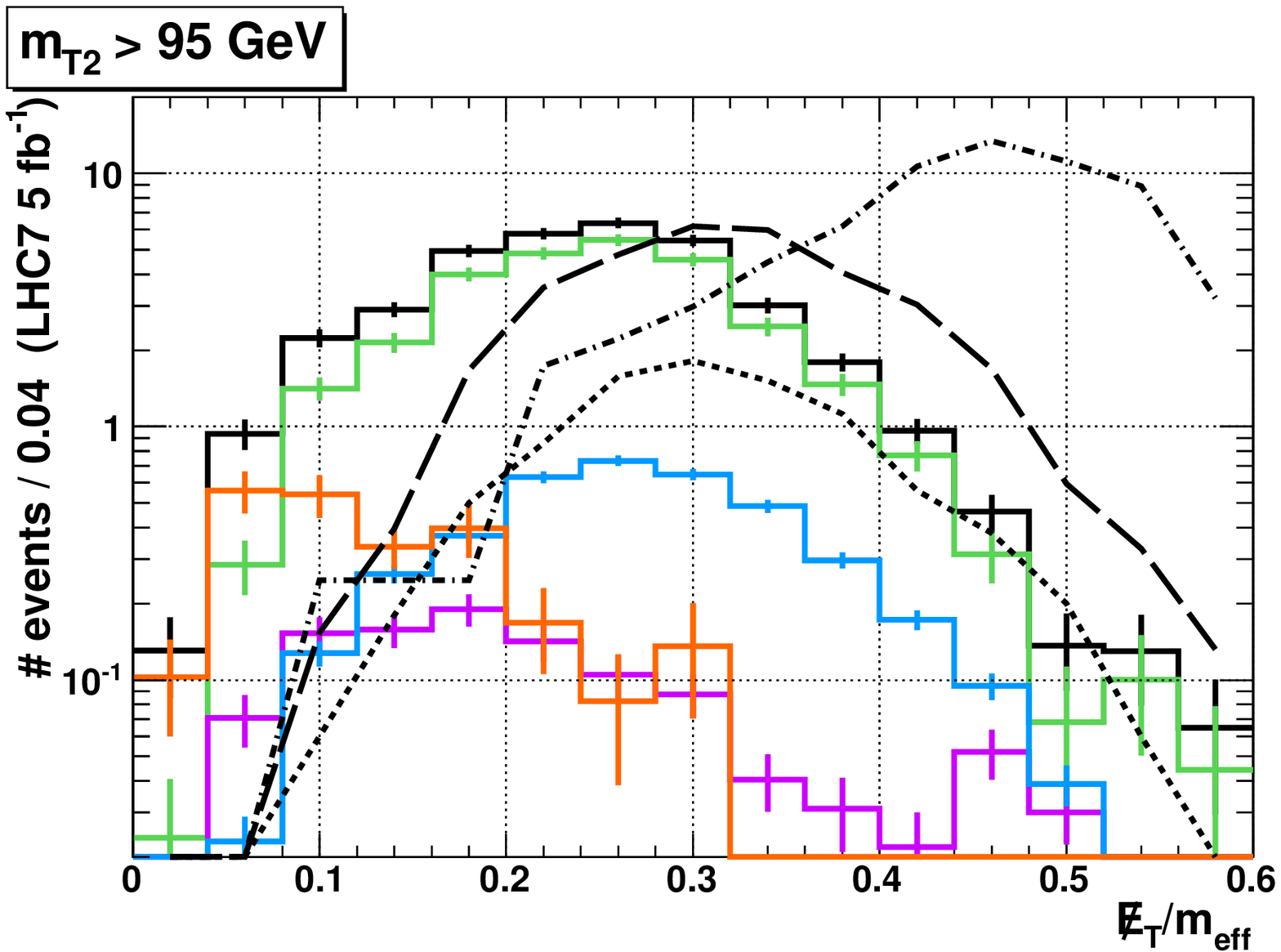}
   \includegraphics[width=3.15in]{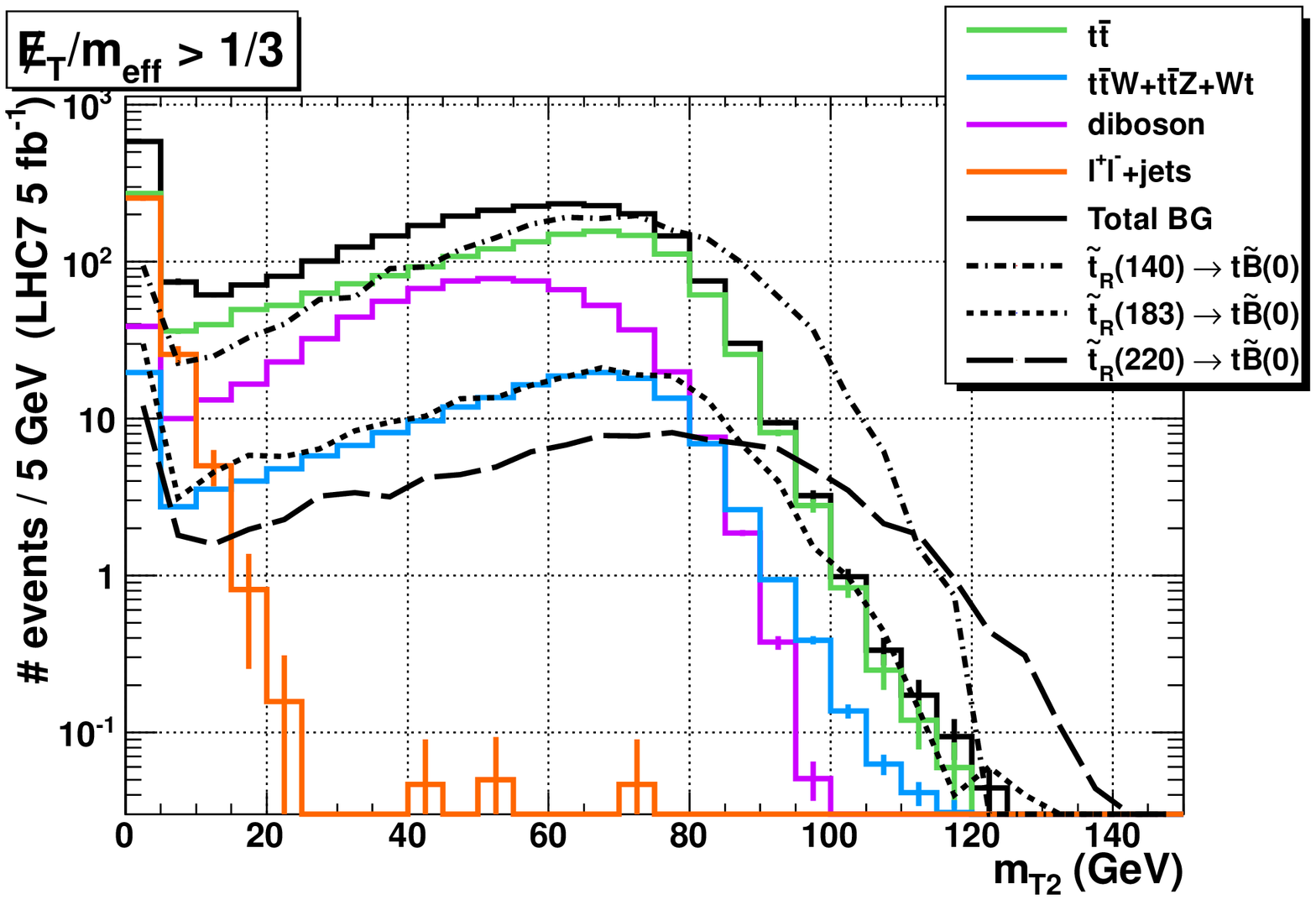}
\end{center}
\caption{$\MET/m_{\rm eff}$ distributions after application of our final $m_{T2} > 95$~GeV cut (left) and $m_{T2}$ distributions after application of our final $\MET/m_{\rm eff} > 1/3$ cut (right), with 5~fb$^{-1}$ at the 7~TeV LHC.  We show our background predictions:  $t\bar t$ (green), single-top and $t\bar tW/Z$ (blue), electroweak diboson (purple), $l^+l^-$+jets (orange), and total background (black solid).  We also show three signal points, corresponding to a massless bino-like neutralino LSP and right-handed stops with masses of 140~GeV (dot-dashed curve), 183~GeV (dotted curve), and 220~GeV (dashed curve).  (Error bars are Monte Carlo statistics.)}
\label{fig:final-dist}
\end{figure}

Having relaxed the jet selection cuts of the ATLAS analysis, we already gain some improvement in $S/B$ at high $m_{T2}$, as seen in Fig.~\ref{fig:baseline-dist}.  However, it is possible to make further gains.  A commonly-used variable for SUSY searches, though not applied by ATLAS for dileptonic stops, is the ratio $\MET/m_{\rm eff}$ (or $\MET/S_{T}$ in CMS terminology).  Here $m_{\rm eff}$ ($S_T$) represents the scalar-summed $p_T$ of all reconstructed objects in the event, including jets, leptons, and $\MET$.\footnote{Other common variations are $\MET/\sqrt{m_{\rm eff}}$ or $\MET/\sqrt{H_T}$, the latter using the scalar $p_T$ sum over jets only.  We find slightly better results with $\MET/m_{\rm eff}$ rather than $\MET/\sqrt{m_{\rm eff}}$, but the choice of exactly how to represent this variable is not crucial.  However, $\MET/\sqrt{H_T}$ is potentially dangerous because our very loose jet requirements can lead to a vanishing denominator.}  We also show in Fig.~\ref{fig:baseline-dist} the distribution of $\MET/m_{\rm eff}$ with our baseline selection.  Some signal and background separation is evident there, especially for the 140~GeV stop.  But in Fig.~\ref{fig:final-dist}, we can see what happens after a 95~GeV $m_{T2}$ cut.  Then, even the difficult 183~GeV stop can be seen to achieve $S/B \sim 1$.  This means that stop events tend to achieve large $m_{T2}$ values with softer leptons and harder $\MET$ relative to backgrounds such as $t\bar t$.  At the same time, we see that non-top backgrounds almost completely fall away, as these usually rely on smaller $\MET$ in association with very hard dilepton systems.

In order to capitalize on the good separation in $\MET/m_{\rm eff}$ at high-$m_{T2}$, we define our final signal region cut as $\MET/m_{\rm eff} > 1/3$ and $m_{T2} > 95$~GeV.  We can get a complementary picture by looking at the $m_{T2}$ distribution after application of this $\MET/m_{\rm eff}$ cut,  as is also shown in Fig.~\ref{fig:final-dist}.

Our complete event selection is then summarized as follows:
\begin{itemize}
\item  Exactly two opposite-sign isolated leptons ($l = e$ or $\mu$) satisfying $p_T(e) > 25$~GeV, $p_T(\mu) > 20$~GeV, $|\eta(l)| \lsim 2.5$, and $m(l^+,l^-) > 20$~GeV (regardless of flavors).
\item  In same-flavor (SF) events, a $Z$-veto ($m(l^+,l^-) \neq [71,111]$~GeV) and at least one $b$-tagged jet.
\item  For the final signal region selection, events with $\MET/m_{\rm eff} > 1/3$ and $m_{T2} > 95$~GeV.
\end{itemize}
For 5~fb$^{-1}$ at LHC7, the dominant backgrounds are then $t\bar t$ (4.0 events), $tW$ (0.6 event), $t\bar t W/Z$ (0.3 event), and $ZZ^*$ (0.2 event).  For 25~fb$^{-1}$ at LHC8, the backgrounds all increase by approximately an order of magnitude.

Understanding the systematic errors on these background predictions would require much more sophisticated detector modeling than what we have available.  To proceed, we simply make a reasonable guess, which at a minimum will give some indication of the level at which these errors might affect the search sensitivity.  While the nominal ATLAS analysis estimates an $O$(1) uncertainty for the $t\bar t$ background at $m_{T2} > 120$~GeV, we expect that the errors on our own version will be smaller.  In the original conference note and on the analysis website~\cite{ATLASdilepNote}, ATLAS reports the error on the background prediction in the range $m_{T2} = [100,120]$~GeV as 15\% for same-flavor and 40\% for different-flavor.  For our own 95~GeV cut, we expect even smaller uncertainty, since ATLAS shows a general trend of decreasing error for decreasing $m_{T2}$ (as can be seen in Fig.~2 of~\cite{Aad:2012uu}).  We also expect the errors to decrease with larger background Monte Carlo samples passed through the full detector simulation, as simulation statistics are clearly still an important limiting factor for the ATLAS analysis.  We therefore take a somewhat optimistic stance, and for our own analysis pick the smaller of the two errors reported for the $[100,120]$~GeV bin, namely 15\%.  In the next section, we will also present some of our results assuming the more pessimistic 40\% choice.

Of course, our analysis uses different selection cuts before cutting on $m_{T2}$, essentially swapping a demand of two hard jets for a demand of somewhat large $\MET/m_{\rm eff}$.  We have no way of assessing whether this significantly changes the error accounting.  However, the $\MET/m_{\rm eff}$ cut is not very harsh and is defined in a manner that is quite inclusive over event kinematics.  We therefore do not expect this modification to pose a significant barrier to achieving a reasonable level of uncertainty.

One place where we do have a certain degree of control over uncertainties is the theoretical error on the background prediction.  We have checked that independent $O$(1) variations on the renormalization and factorization scales of our $t\bar t$ simulations, generated with {\tt MC@NLO}, lead to variations in our predictions which are at the level of our Monte Carlo statistics.  This is roughly 10\% for events passing our final cuts.

\section{Results}
\label{sec:results}

Having outlined the details of our search strategy optimized for light stops in the compressed regime, we now estimate the sensitivities that can be obtained with this analysis.  To make contact with the existing 2011 ATLAS study, which placed exclusions for $\stop_R \to t\Bino$, we first describe how those exclusions might be expanded.  We also show how they extend to the case of $\stop_R \to t\gold$, to which there is even better sensitivity.  We then show the prospects for the full 2012 data sample at 8~TeV.  We perform our analysis for all four limiting models described in section~\ref{sec:spin}.  For the neutralino LSP cases, we also show a reinterpretation of ATLAS's $\stop \to b \tilde\chi^+_1$ search, which capitalizes on low-$p_T$ dileptons and further expands the possible coverage in the $(m_{\stop},m_{\neu})$ mass plane for very light stops.  Finally we turn to the question of how the size of systematic errors may impact our $t_R \to t\Bino$ results, and we make projections for what may be achievable in the future.  We consider both a cut-and-count style analysis after a high statistics run at 14~TeV, and the possibility of a precision shape measurement of the $m_{T2}$ spectrum.

Below, we base our main statistical results on counting experiments using the $CL_S$ procedure.  As discussed above, we define our signal region (SR) using a set of baseline selection cuts as well as a final selection with $\MET/m_{\rm eff}>1/3$ and $m_{T2}>95$~GeV.  Unless stated otherwise, we assume a 15\% systematic error on the background prediction in this region.

Analyses of this type also usually normalize backgrounds using control regions (CR).  This would not be a necessary step for a simple estimate such as ours, but we include it to avoid misleading results.  For very compressed spectra or unfavorable chirality choices, the $\MET/m_{\rm eff}$ and $m_{T2}$ distributions can look very top-like, and a realistic analysis can potentially normalize away the signal.  The actual choice of a reasonable CR would ultimately be determined by the experimentalists.  We simply take as our CR sample the entire set of events satisfying our baseline selection and inverting our final selection: $\MET/m_{\rm eff} < 1/3$ or $m_{T2} < 95$~GeV.  (We neglect the non-top backgrounds such as diboson and $l^+l^-$+jets in this construction, as these can be separately normalized using alternative control regions, and are highly subdominant in the SR.)  Effectively, our nominal statistical procedure corresponds to a shape analysis with two bins.  We provide complete details in appendix~\ref{sec:statistics}.

\subsection{Prospects for 2011 data}

\begin{figure}
   \includegraphics[width=3.15in]{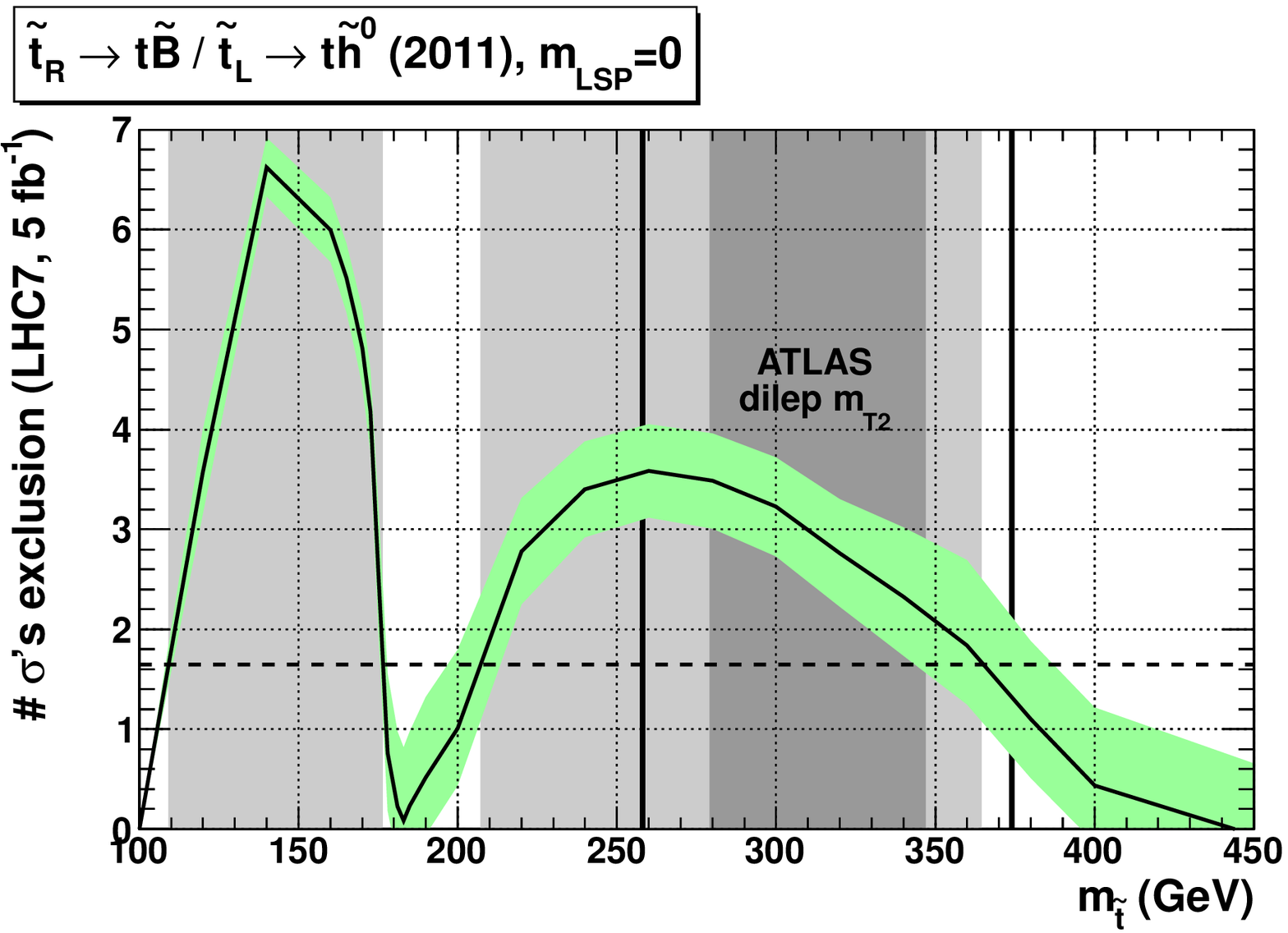}
   \includegraphics[width=3.15in]{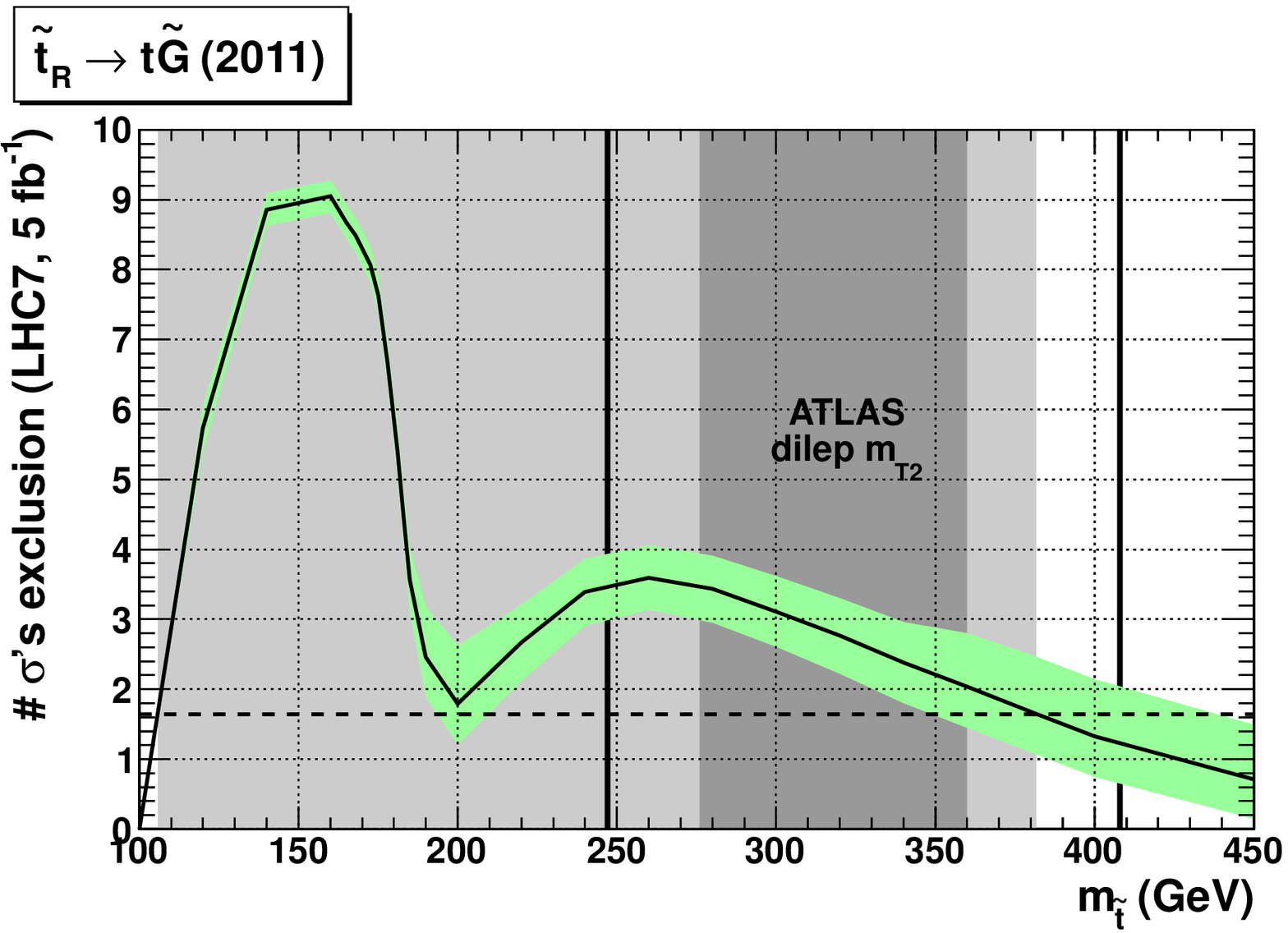}
\caption{Estimated exclusion level for $\stop_R \to t\Bino$/$\stop_L \to t\higgsino$ with massless LSP (left) and $\stop_R \to t\gold$ (right), assuming a 2011-like data sample of 5~fb$^{-1}$ at 7~TeV.  The black curve and green band show the median and $\pm1\sigma$ quantile exclusions, with 15\% systematic errors on the background.  The dashed horizontal line indicates 95\% $CL_S$ exclusion.  The light gray shaded region is where our median exclusion is better than 95\%.  The dark grey shaded region and vertical black lines are ATLAS's observed and expected exclusions from their dileptonic $m_{T2}$ search.}   \label{fig:excl7}
\end{figure}

In Fig.~\ref{fig:excl7}, we show the $CL_S$ exclusion level estimated for the 2011 LHC data set of 5~fb$^{-1}$ at 7~TeV, for the two scenarios $\stop_R \to t\Bino$ (equivalent to $\stop_L \to t\higgsino$) with massless LSP and $\stop_R \to t\gold$.  The $CL_S$ exclusion is translated into an effective number of one-sided $\sigma$'s, with 95\% exclusion corresponding to 1.64$\sigma$.  In addition to the median exclusion expected with a background-only sample, we show the the usual ``green band'' of 1$\sigma$ statistical uncertainty on the exclusion prediction.  The plots also show the ATLAS 2011 exclusion using dileptonic $m_{T2}$, which we have re-evaluated for the gravitino case and for spectra with off-shell decays.

For the $\stop_R \to t\Bino$ case, we can see that the potential exclusion range is greatly expanded on the low end, pushing into the compressed region and ruling out a large swath of the region with $m_{\stop} < m_t$.  A gap nonetheless remains between 175~GeV and 210~GeV.  We will determine the fate of this gap in the next subsection.

For $\stop_R \to t\gold$, the prospects are even better, owing to the typically larger energies carried by the gravitino.  Our median prediction is that the entire mass range $m_{\stop} = [105,380]$~GeV can already be ruled out.  However, the weakest point, at $m_{\stop} = 200$~GeV, is just barely closed off by the median exclusion.  It is possible that statistical fluctuations would lead to a weaker observed exclusion, leaving a small gap in the mass coverage.

\subsection{Prospects for 2012 data}
\label{sec:results2012}

The larger luminosity and energy of the 2012 LHC run will greatly increase the sensitivity of the dileptonic $m_{T2}$ search.  For the results presented below, we assume 25~fb$^{-1}$ of data collected at 8~TeV.

\begin{figure}
\begin{center}
   \includegraphics[width=3.15in]{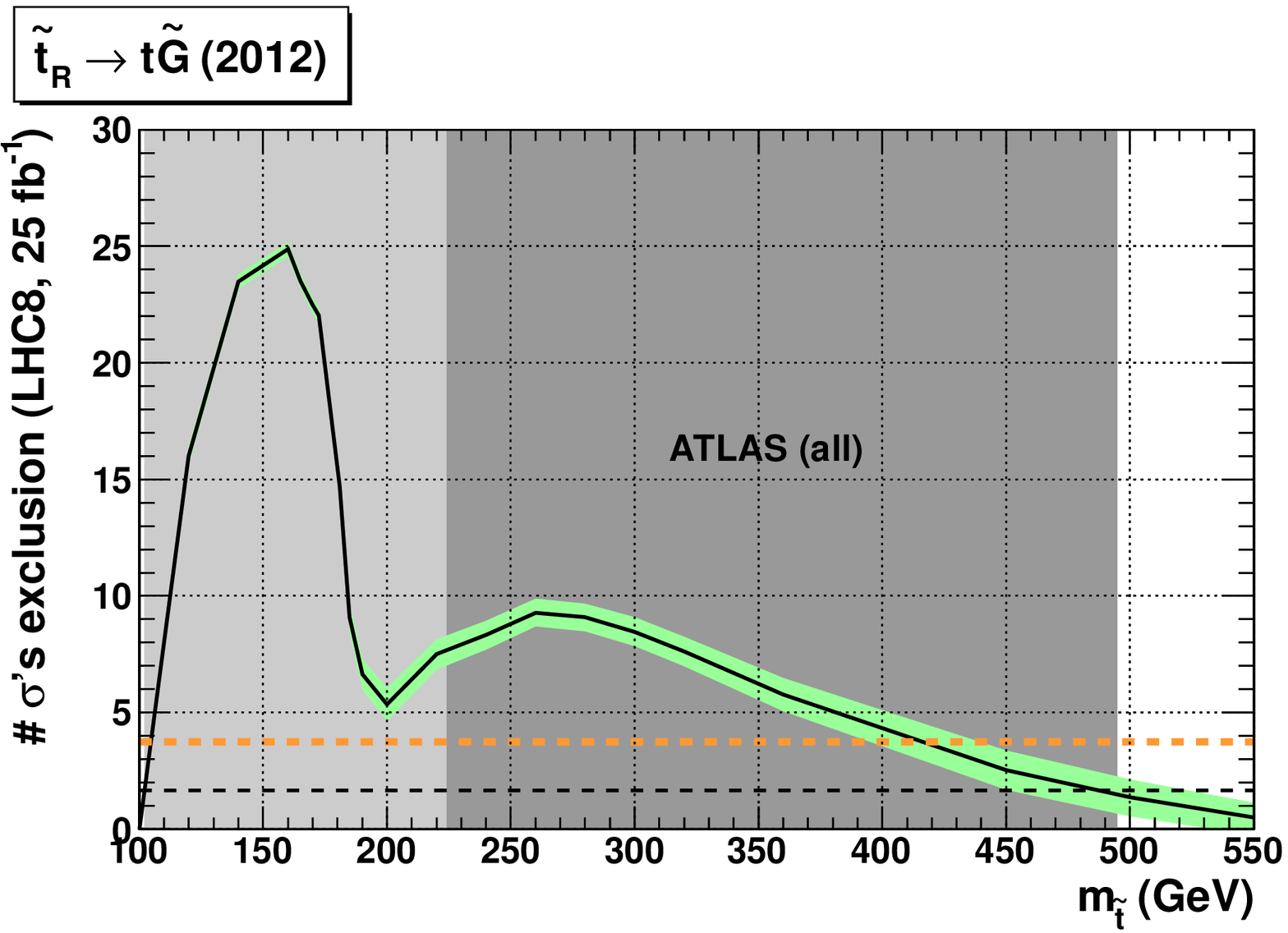}
   \includegraphics[width=3.15in]{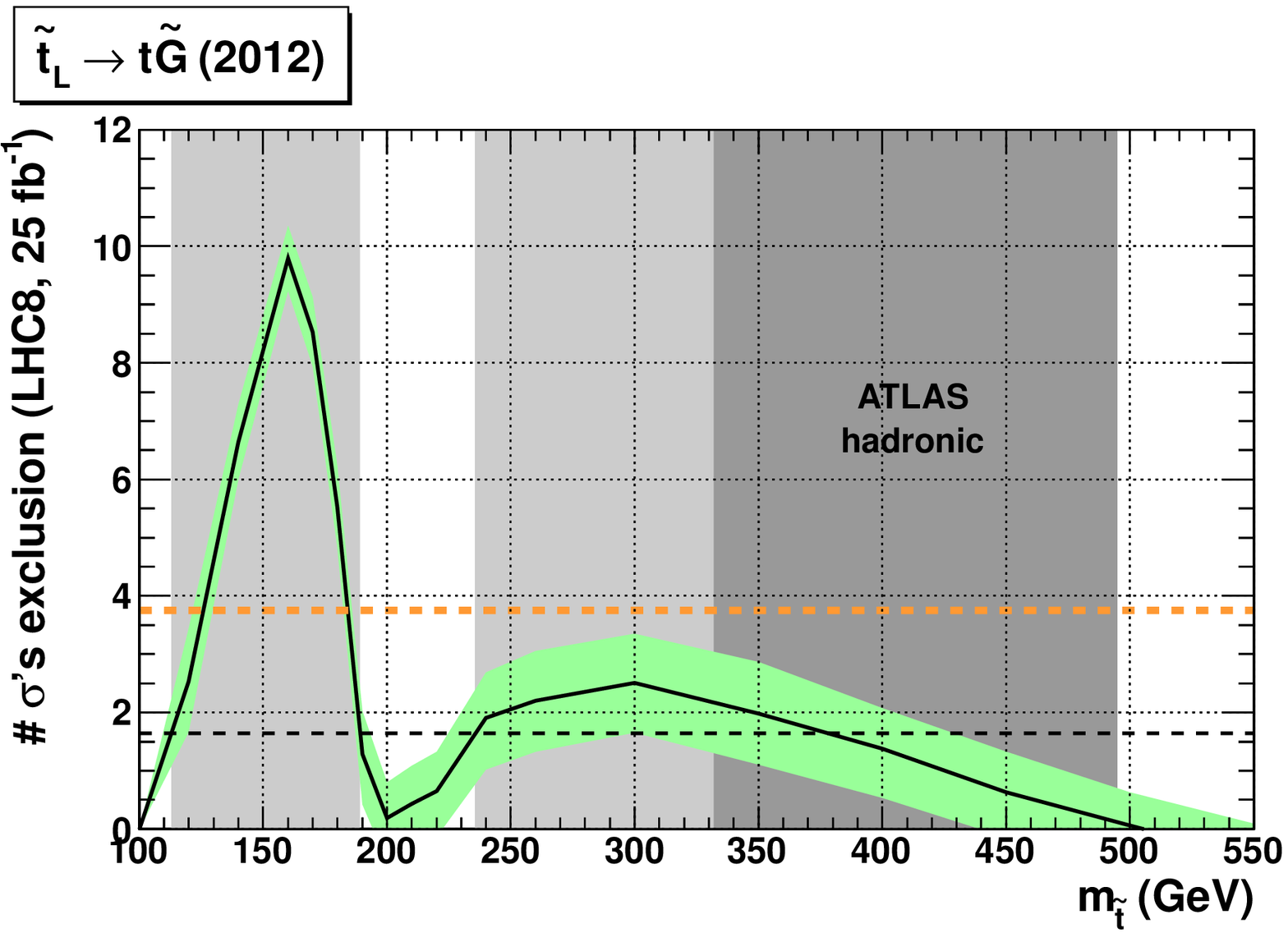}
\end{center}
\caption{Estimated exclusion level for $\stop_R \to t\gold$ (left) and $\stop_L \to t\gold$ (right), assuming a 2012-like data sample of 25~fb$^{-1}$ at 8~TeV.  The black curve and green band show the median and $\pm1\sigma$ quantile exclusions, with 15\% systematic errors on the background.  The dashed horizontal black line indicates 95\% $CL_S$ exclusion, and the dashed horizontal orange line indicates the approximate equivalent of ``5$\sigma$'' discovery level.  The light gray shaded region is where our median exclusion is better than 95\%.  On the left plot, the dark gray shaded region indicates the complete range of ATLAS exclusions: dileptonic, $l$+jets, and all-hadronic.  (The left edge is controlled by $l$+jets, which we have not re-interpreted for a gravitino LSP.  We expect the true exclusion to be stronger.)  On the right plot, the dark gray shaded region indicates the ATLAS all-hadronic exclusion, which is likely their only search unaffected by the top quark's spin.}   \label{fig:excl8grav}
\end{figure}

We begin with the gravitino LSP.  In Fig.~\ref{fig:excl8grav}, we show how the coverage will evolve for $\stop_R$, and now also include results for $\stop_L$.  The former search becomes capable of cleanly excluding stops between 100~GeV and 490~GeV, with no gaps.  Indeed, most of the range not already excluded by ATLAS stop searches would exceed discovery-level significance, though some of this region might be independently excluded by non-dedicated SUSY searches~\cite{Kats:2011it,Kats:2011qh}.  In the $\stop_L \to t\gold$ search, we clearly see the degrading effects of left-handed top quark polarization.  The coverage is much weaker over the entire range (note the change in vertical scale).  We nonetheless predict exclusion-level sensitivity in the ranges $m_{\stop} = [110,190]$~GeV and $[235,380]$~GeV, and a region with discovery-level sensitivity centered at 160~GeV.  While a gap in coverage remains, we expect this to close off as even more data comes in at the upgraded LHC.\footnote{More generally, mixed stops may or may not exhibit this gap.  For the weakest mass point, $m_{\stop} \simeq 200$~GeV, scanning over stop mixing angles, we estimate that the borderline case is a mostly-$\stop_L$ with $|\theta_{\stop}| \simeq 40^\circ$.}  We also emphasize that our own analysis has not been separately optimized for this case, and a more careful event selection might yield better results.  In particular, $\stop_L \to t \gold$ could benefit from a harder $m_{T2}$ cut.

\begin{figure}
\begin{center}
   \includegraphics[width=5in]{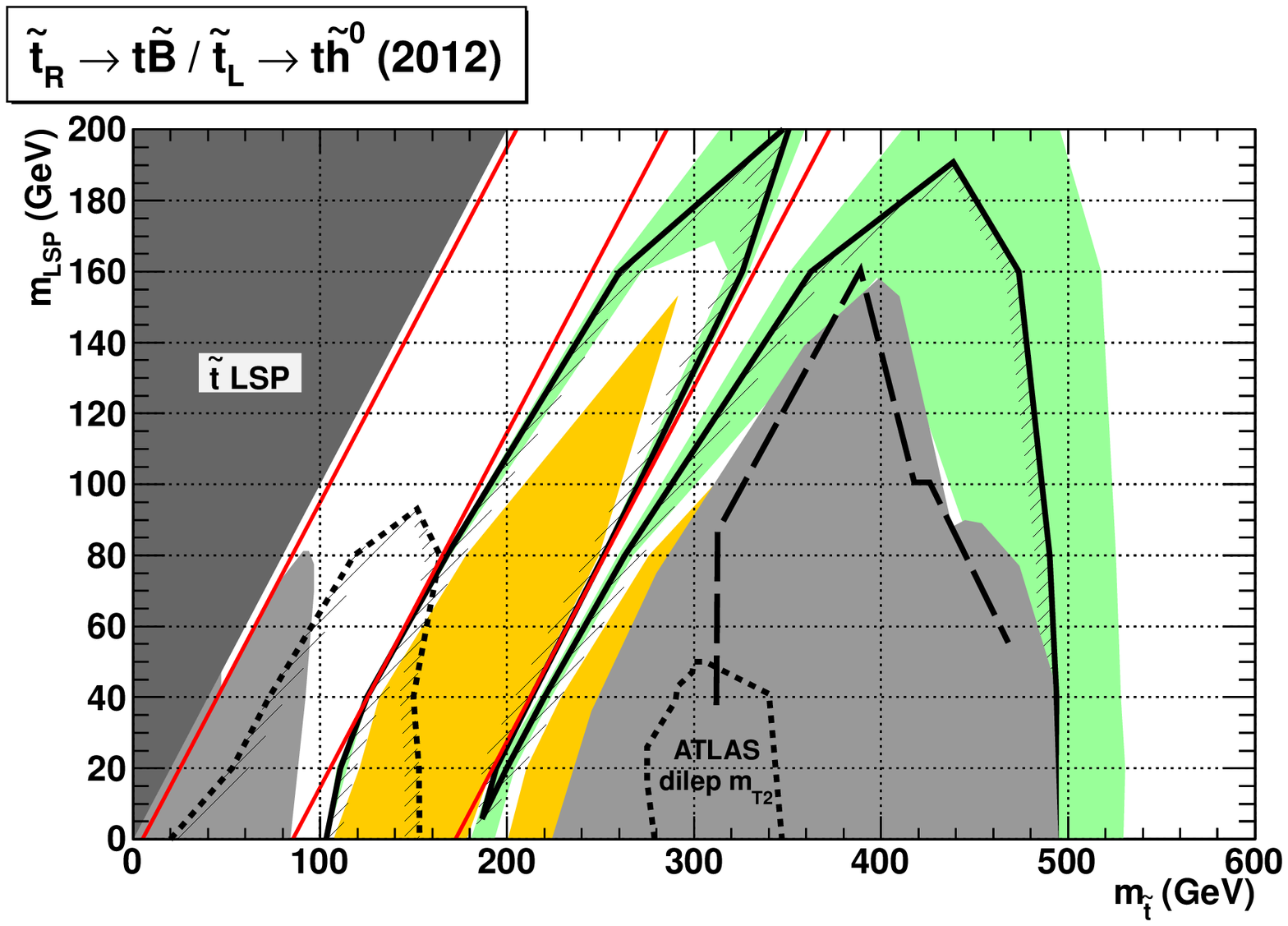}
\end{center}
\caption{Estimated 95\% exclusion region and 5$\sigma$ discovery region for $\stop_R \to t\Bino$/$\stop_L \to t\higgsino$, assuming a 2012-like data sample of 25~fb$^{-1}$ at 8~TeV.  Our median exclusion boundary is represented by the solid black line with hash marks, and the $\pm1\sigma$ quantile boundaries define the green band.  Discoverable regions are shaded orange.  We also include various existing experimental constraints.  Low-mass LEP exclusions and the complete set of high-mass ATLAS exclusions are shaded light gray, with the ATLAS dileptonic $m_{T2}$ region bordered by the dotted black line.  The exclusion boundary from CMS all-hadronic searches (inclusive razor, $b$-tagged razor, and $\alpha_T$) is indicated by the dashed black line.  The dotted black line with hashes shows the exclusion possible from the ATLAS low-$p_T$ dilepton search for $\stop \to b\tilde\chi^+_1$.  Red lines indicate the boundaries between the different $N$-body kinematic regions.}   \label{fig:excl8bino}
\end{figure}

\begin{figure}
\begin{center}
   \includegraphics[width=2.5in]{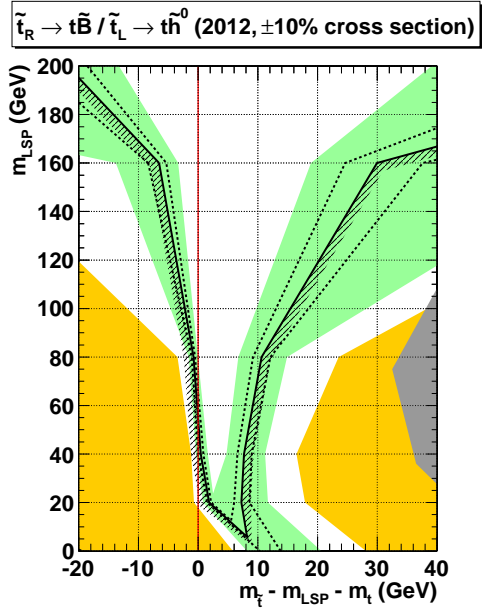}
\end{center}
\caption{A zoom-in of the compressed region of Fig.~\ref{fig:excl8bino}, using shifted coordinates.  The dotted lines show the modification to the median exclusion boundary under $\pm$10\% variations of the signal cross section.}   \label{fig:zoom}
\end{figure}

For $\stop_R \to t\Bino$ and $\stop_L\to t\higgsino$, we extend our results into the full mass plane, shown in Fig.~\ref{fig:excl8bino}.  Compared to the original 2011 ATLAS dileptonic search, the possible range of coverage has expanded dramatically, encapsulating a large portion of both the 2-body and 3-body decay regions.  We achieve discovery-level significance over broad parts of the plane, including an area between the current exclusion and the compressed region at the 2-body/3-body interface.  Our exclusion-level sensitivity extends deep into this region, just barely closing the gap for a massless $\neu$ and squeezing it down to $O$(10)~GeV width for higher masses.  We show a more detailed zoom-in on the compressed region in Fig.~\ref{fig:zoom}.  There, we also show the effect of $\pm10\%$ variations in the signal cross section, which give a rough indication of theoretical errors.  We see that the size of the gap is not very sensitive to these cross section variations.  As we will see in the next subsection, its extent is however highly dependent on systematic errors.  

We also indicate in Fig.~\ref{fig:excl8bino} the exclusions from the full complement of ATLAS searches (dileptonic~\cite{Aad:2012uu}, $l$+jets~\cite{:2012ar}, and all-hadronic~\cite{:2012si}), the various CMS all-hadronic searches~\cite{Chatrchyan:2012wa,CMSrazor,CMSbtagRazor}, and LEP limits from a direct 3-body/4-body stop search by ALEPH\footnote{In particular, we use the results of ALEPH's search for $\stop \to bl\nu\neu$ via an off-shell chargino.  While we assume that a different diagram topology dominates, namely decay via off-shell top quark, we have found that the two cases have very similar decay kinematics.  Chirality effects in these searches should also be less pronounced, since they are much more inclusive.  The results presented here represent an average between the (very similar) constraints placed on pure a left-handed stop and a $\theta_{\stop} = 56^\circ$, mostly-right-handed stop with vanishing $Z$ coupling.}~\cite{Heister:2002hp} and from $Z$ width constraints~\cite{LEPEWWG}.  We supplement these limits with a re-interpretation of ATLAS's 2011 search for low-mass stops decaying as $\stop\to b\tilde\chi^+_1 \to b(W^+\neu)$, which capitalizes on low-$p_T$ dileptons~\cite{Aad:2012tx}.  This makes a substantive impact in the fully 4-body region, where our own strategy no longer works because the $W$ bosons are off-shell.  We have also studied the dileptonic portion of ATLAS's low-mass stop search based on the cluster transverse mass variable~\cite{Aad:2012yr}, but obtain somewhat weaker results.  It would be interesting to make a more comprehensive study of these searches, and to better optimize them for direct stop decays without assuming intermediate on-shell charginos.

\begin{figure}
\begin{center}
   \includegraphics[width=5in]{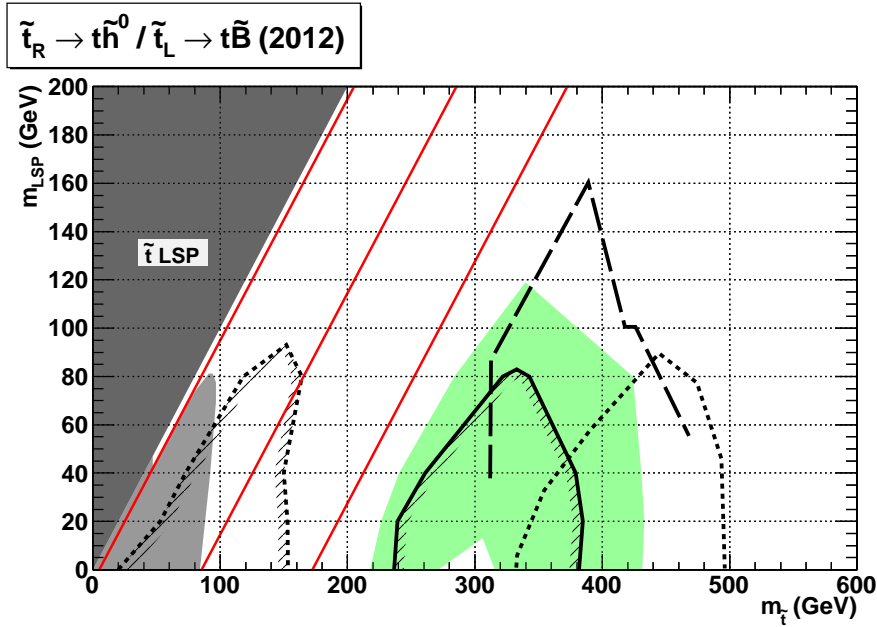}
\end{center}
\caption{Estimated 95\% exclusion region for $\stop_R \to t\higgsino$/$\stop_L \to t\Bino$, assuming a 2012-like data sample of 25~fb$^{-1}$ at 8~TeV.  Our median exclusion boundary is represented by the solid black line with hash marks, and the $\pm1\sigma$ quantile boundaries define the green band.  We also include various existing experimental constraints.  Low-mass LEP exclusions are shaded light gray.  The ATLAS all-hadronic search boundary is indicated by the dotted black line.  The exclusion boundary from CMS all-hadronic searches (inclusive razor, $b$-tagged razor, and $\alpha_T$) is indicated by the dashed black line.  The dotted black line with hashes shows the exclusion possible from the ATLAS low-$p_T$ dilepton search for $\stop \to b\tilde\chi^+_1$.  Red lines indicate the boundaries between the different $N$-body kinematic regions.}   \label{fig:excl8hu}
\end{figure}

By comparison, the prospects for $\stop_R \to t\higgsino$ and $\stop_L \to t\Bino$ are much less optimistic.  In the 2011 data, neither our search nor ATLAS's dileptonic $m_{T2}$ search are capable of placing any limits.  The situation in 2012 is shown in Fig.~\ref{fig:excl8hu}.  With rather sizeable uncertainty, we can exclude stops in the range $[240,380]$~GeV given a massless LSP.  No region passes discovery-level significance.  We note that a similar, though perhaps somewhat less severe degrading of sensitivity would also occur in $l$+jets searches based on $m_T(l,\MET)$, as illustrated above in Fig.~\ref{fig:left-vs-right}.  Consequently, we do not show existing LHC exclusions for dileptonic or $l$+jets searches for these models, but only for all-hadronic searches.  However, the same LEP limits should continue to apply, as well as those from the ATLAS low-$p_T$ dilepton and low cluster transverse mass searches.  It would clearly be interesting here to see how the all-hadronic searches perform for off-shell decays, as well as how far they can be pushed into the compressed region using precision shape measurements (as suggested by~\cite{Alves:2012ft}).  We also comment that the dileptonic search proposed by~\cite{Han:2012fw}, which focuses on distortions of $t\bar t$ spin correlation and rapidity distributions, should still apply.

\subsection{Effects of systematic errors and future outlook}
\label{sec:outlook}

\begin{figure}
\begin{center}
   \includegraphics[width=2.5in]{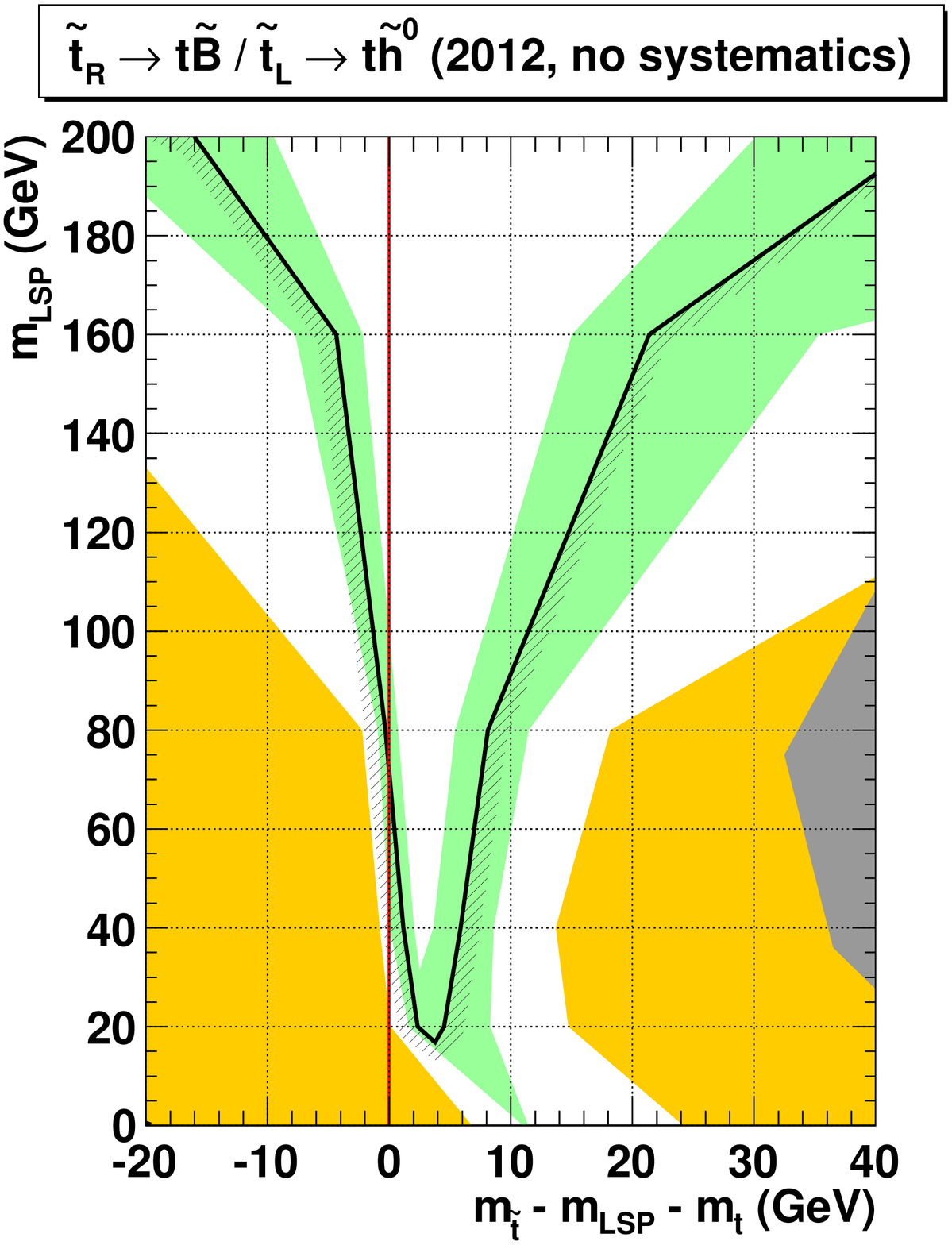}
   \includegraphics[width=2.5in]{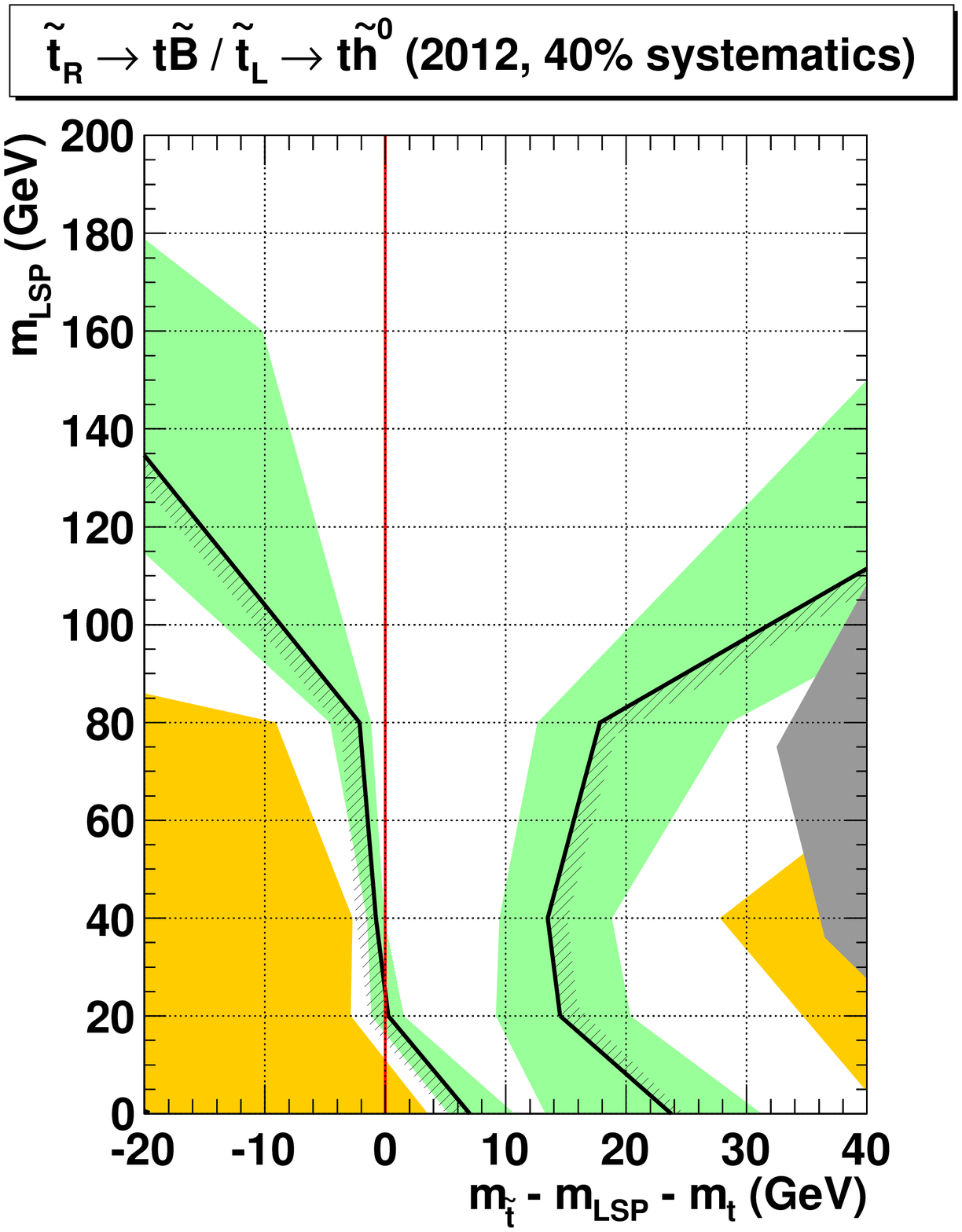}
\end{center}
\caption{As in Fig.~\ref{fig:zoom}, a zoom-in of the compressed region of Fig.~\ref{fig:excl8bino}, using shifted coordinates.  The systematic errors have been changed from 15\% to 0\% (left) and 40\% (right).}   \label{fig:syst-bg2}
\end{figure}

All of our results so far assume a 15\% systematic error on the background.  Over large portions of parameter space, the precise choice of systematic error does not significantly alter our estimates of exclusion or discovery potential.  However, we briefly address the impact of these errors in the difficult compressed region for $\stop_R \to t\Bino$/$\stop_L \to t\higgsino$.  In Fig.~\ref{fig:syst-bg2}, we show how the boundaries of the exclusion change if we either optimistically assume vanishing systematics, or pessimistically assume 40\% systematics.  In the former case, the low-mass LSP is much more cleanly closed off, with median exclusion extending up to about 17~GeV for any stop mass in this range.  The remaining sensitivity gap is also generally narrower.  In the 40\% systematics case, the gap instead becomes much broader, already 15~GeV wide for a massless LSP.  Clearly, maintaining good control over the systematic errors will be crucial to narrowing the gap as much as possible, and having a good understanding of the size of those errors will be necessary to reliably delineate the exclusion boundary in this region. 

\begin{figure}
\begin{center}
   \includegraphics[width=2.5in]{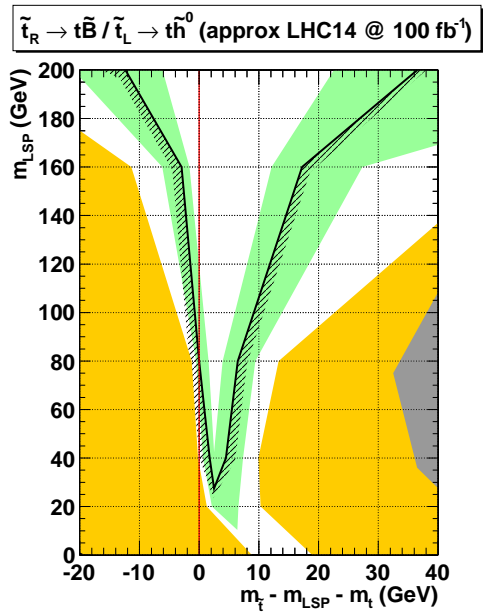}
\end{center}
\caption{As in Fig.~\ref{fig:zoom}, a zoom-in of the compressed region of Fig.~\ref{fig:excl8bino}, using shifted coordinates.  Systematic errors are again 15\%, but the effective luminosity has been increased by a factor of 15, to roughly extrapolate to what is possible with 100~fb$^{-1}$ at 14~TeV using our simple cut-and-count search.}   \label{fig:14TeV}
\end{figure}

We can also consider the improvements in sensitivity that may be achieved with a high statistics dataset corresponding to 100~fb$^{-1}$ at 14~TeV. To estimate this, we scale up the event counts used in computing the 2012 results by a factor of 15, to account for the enhanced luminosity as well as the increased signal and background cross sections. The projected sensitivity in the compressed region with these assumptions is shown in Fig.~\ref{fig:14TeV}.  It is evident from this plot that statistics were still a major limiting factor for the 2012 analysis.  The exclusion gap for a massless LSP can now be closed entirely.  (At this point we instead have a {\it discovery} gap, of roughly 10~GeV width.)  Median exclusion fully covers LSP's up to 28~GeV. We emphasize again that our search has not been re-optimized for these variations, thus higher sensitivity may well be achievable.

Considering the power of a simple cut-and-count search using the dileptonic $m_{T2}$ variable, a straightforward way to further enhance sensitivity would be to measure the detailed shape of the $m_{T2}$ spectrum.  Our own initial studies with {\tt MC@NLO} indicate that the shape is quite stable to $O(1)$ renormalization and factorization scale variations, as might be expected given the highly inclusive nature of our kinematic criteria.  A more comprehensive study with higher statistics, also folding in NLO corrections to the top decay, would be interesting to pursue.  As a first estimate of the potential of a shape-based search, we neglect all possible theoretical and experimental systematic errors.  We divide the $m_{T2}$ distribution, after the $\MET/m_{\rm eff}$ cut, into 5~GeV bins from 20~GeV to 125~GeV and a single overflow bin for events above 125~GeV.  We include all backgrounds, though the top-like background ($t\bar t$, $tW$, and $t\bar t W/Z$) is by far the most important.  In constructing the signal+background shape, we add the stop signal and the top-like background, which tend to have similar shapes at low $m_{T2}$, and rescale to the top-like background's original normalization.  To build up probability density functions for background-only and signal+background hypotheses, we compute log-likelihood ratios over an ensemble of pseudo-experiments, as described in appendix~\ref{sec:statistics}.  A scan over the weakest $(m_{\stop},m_{\neu})$ points from our cut-and-count search reveals major improvements.  For the 2012 analysis, the gap in median exclusion completely closes for $m_{\neu} \lsim 45$~GeV.  Extrapolating to 100~fb$^{-1}$ at 14~TeV, complete exclusion coverage extends up to $m_{\neu} \simeq 200$~GeV.  Clearly, it will be worthwhile to understand whether the $m_{T2}$ spectrum can be predicted accurately enough to take advantage of such an analysis.

In the case of a gravitino LSP, we have shown that excellent coverage can be obtained with the 2012 dataset using a simple cut-and-count search, if the stop is right-handed.  But a 45~GeV gap remained for left-handed stops.  Projecting to 14~TeV running with our nominal analysis, we expect this gap to shrink significantly, down to 5--10~GeV, and could likely be closed with harder cuts.  The shape-based analysis for 2012 would also already just barely close the gap.

Finally, $\stop_R \to t\higgsino$ and $\stop_L \to t\Bino$ will continue to be challenging, even at high statistics or with more powerful analysis procedures.  For compressed and off-shell decays, the bulk of the signal $m_{T2}$ shape can actually become {\it softer} than the background $t\bar t$ shape, though the high-$m_{T2}$ tail can still be broader.  Harder cuts at the 14~TeV LHC could potentially pick up the latter feature.  In principle, the shape-based analysis can pick up both of these features, extending the 2012 exclusion for massless higgsinos down from 240~GeV to 215~GeV, and to masses well below 200~GeV for 14~TeV.  Nonetheless, other search strategies may yield more immediate gains.

\section{Summary and Conclusions}
\label{sec:conclusions}

The LHC offers our best chance to comprehensively search for the supersymmetric top partners that may be responsible for taming fine-tuning in the Higgs sector.  Given the importance of this endeavor, and the current lack of signals in the most straightforward searches, it is becoming increasingly clear that more aggressive strategies must be employed.  In this paper, we have explored a very promising option for expanding direct stop searches into the poorly-covered compressed and off-shell decay regions, utilizing a simple modification to ATLAS's existing dileptonic $m_{T2}$ search.

Dileptonic $m_{T2}$ is perhaps the cleanest variable available for a stop search, as the high-$m_{T2}$ background is almost entirely dileptonic $t\bar t$, and is very sharply falling beyond $m_W$.  Primarily, we suggest lowering the final $m_{T2}$ cut used by ATLAS, which both improves sensitivity to lighter stops and reduces systematic errors.  We also propose that this search be run in a more inclusive manner, essentially placing no explicit cuts on jets except for a $b$-tag in the same-flavor channel.  Non-top backgrounds can otherwise be almost entirely suppressed by a cut on $\MET/m_{\rm eff}$, which also helps significantly to separate stop pairs from top pairs.  The final separation is good enough that a simple cut-and-count analysis with $O(10\%)$ systematic errors is capable of making major gains in light stop coverage.

We further propose that this search and other stop searches be applied to a broader class of models.  The typical assumption of $\stop_R \to t\Bino$ is nominally well-motivated due to its simplicity, but many other possibilities exist.  In particular, we have identified four limiting cases with distinct exclusion or discovery prospects, encompassing six different model choices.  These cover contributions from both stop chiralities and a broad range of possible LSP's, including those with gravitinos or light singlinos.

The most promising case for the dileptonic $m_{T2}$ search is a right-handed stop decaying into a gravitino LSP, which can nearly be fully excluded at low mass with no gaps using only the 2011 dataset, and will be definitively excluded or discovered in 2012.  Left-handed stops are more difficult, due to a softer $m_{T2}$ spectrum arising from top polarization effects.  Even after the 2012 run, there may be a roughly 50~GeV wide gap remaining in exclusion sensitivity.

The popular $\stop_R \to t\Bino$ case has identical phenomenology to $\stop_L \to t\higgsino$. 2011 data already allows exclusion-level sensitivity down to 210~GeV stops with a massless LSP, with a large swath of off-shell decays also excluded up to just beyond $m_t$. The 2012 data will further expand this range, essentially closing off light stops of any mass with a very light LSP, and leaving open only an $O$(10~GeV) wide corridor in the compressed region with more massive LSP's.  At the same time, new regions with discovery-level sensitivity open up, including 3-body decays with LSP's up to 150~GeV, and an interesting semi-compressed region where the stop undergoes a 2-body decay with only 20$\sim$40~GeV of available kinetic energy.

An alternative set of models features $\stop_R \to t\higgsino$, or equivalently $\stop_L \to t\Bino$.  These tend to be the most difficult to search for using dileptonic $m_{T2}$, as they have the least favorable combination of spin and momentum-scaling effects in their decays.  We estimate that these cases are unobservable in 2011 $m_{T2}$ searches, and only marginally visible in 2012 searches.  We emphasize that similar degrading effects also afflict $l$+jets searches based on $m_T(l,\MET)$, though they should be somewhat less pronounced.  While we expect dileptonic $m_{T2}$ to play a more limited role in searches for these models, it is not clear which alternative search would offer the best sensitivity, or whether the compressed regime can be fully covered for any LSP mass. One of the best remaining options may be the precision measurement of $t\bar t$ spin correlations and rapidity distributions, as proposed by~\cite{Han:2012fw}.

Our work can point the way to possible future improvements and extensions.  An obvious question is whether the small gap in coverage for compressed $\stop_R \to t\Bino$ can ever be fully closed.  Significant progress can certainly be made by the 14~TeV LHC, just exploiting the much higher statistics.  In order to obtain higher sensitivity still, one may attempt a more ambitious $m_{T2}$ shape analysis rather than a cut-and-count approach. Our preliminary results suggest that a shape analysis is capable of closing the exclusion gap for neutralinos up to 45~GeV with 2012 data, and potentially up to 200~GeV with 14~TeV data.  The other models that are difficult to cover with the simple cut-and-count search, because they involve left-handed top quarks, also become much more feasible.

We have seen that while our method is extremely efficient in covering the 3-body stop decay region, it gives no coverage when the $W$ bosons are forced off-shell in the fully 4-body stop decay.  We have indicated that a re-interpretation of the low-mass $\stop \to b\tilde\chi^+_1$ searches by ATLAS already leads to some progress in covering the 4-body region. It would be interesting to explore how far such searches can be pushed.  Conversely we expect that our $m_{T2}$ approach can lead to progress in searching for $\stop \to b\tilde\chi^+_1$ in regions of parameter space where the $W$ boson is on-shell.  More generally, searches for various other SUSY spectra or non-SUSY models can benefit, in particular searches for fermionic top partners.

Finally, we comment that all of our signal modeling has been done at leading-order, supplemented by parton showers and in some cases shower/matrix-element matching.  To conclusively understand what is happening for compressed spectra, where we have seen that the kinematics can be subtle, it would be very useful to have access to more precise predictions for $m_{T2}$ distributions modeled at NLO in both production and decay, and to systematically understand the uncertainties on these predictions.  Automated NLO corrections on the production side should soon be possible with the {\tt MadGolem} program~\cite{GoncalvesNetto:2012yt}.  Kinematic distributions for first and second generation squarks, complete through NLO in production and decay, are also already being studied~\cite{Hollik:2012rc}.

Despite the unprecedented center of mass energy and integrated luminosity of the LHC so far, we are still at the early stages of the LHC era.  While searches will continue to probe ever-increasing mass scales, it is important to continue working to understand whether we are already copiously producing relatively light new particles which are nonetheless buried amidst the huge Standard Model backgrounds.  This may in fact be the case for supersymmetry.  We have shown in this paper that even some of the most difficult supersymmetric spectra are not hopelessly buried, and we have pointed to future improvements that can be achieved using the huge statistics expected over the lifetime of the LHC and with more sophisticated analysis techniques.


\vspace{0.8cm}
Note added:  While this work was being completed, CMS announced their version of the direct stop search using $l$+jets and $m_T(l,\MET)$, utilizing almost 10~fb$^{-1}$ of data at 8~TeV~\cite{CMSstopSemilep}.  A major difference between CMS and ATLAS is that the former uses {\tt PYTHIA6}~\cite{Sjostrand:2006za} for its SUSY simulations, losing all spin effects, whereas the latter uses {\tt Herwig++}~\cite{Bahr:2008pv}, which has a complete treatment of spin.  Indeed, the CMS results are weaker than what might be expected by extrapolating from ATLAS's results, and CMS has pointed out that this is because the top quarks in ATLAS's simulations are polarized.

\bigskip
\bigskip

{ \Large \bf Acknowledgments}

\smallskip \smallskip

\noindent
We thank Matthew Baumgart and Matthew Reece for useful conversations.  The research of CK was supported in part by NSF grant PHY-0969020.  The research of BT was supported by DoE grant DE-FG-02-91ER40676 and by NSF grant PHY-0969510 (LHC Theory Initiative).  CK would like to thank the Center for Theoretical Underground Physics and Related Areas (CETUP* 2012) in South Dakota for its hospitality and for partial support during the completion of this work.  We also thank the Aspen Center for Physics, supported by NSF grant PHY-1066293, where part of this work was completed.

\appendix

\section{Simulation}
\label{sec:simulation}

We generate our $t\bar t$ background samples using the {\tt MC@NLO} event generator~\cite{Frixione:2002ik,Frixione:2003ei}, normalizing the cross section to 165~pb (230~pb) for LHC7 (LHC8) (see, e.g.,~\cite{Kidonakis:2011ca} and references therein).  For all of our other samples, including additional $t\bar t$ samples for cross checks, we use {\tt MadGraph5} {\tt v1.4.7}~\cite{Alwall:2011uj} interfaced with {\tt PYTHIA}~\cite{Sjostrand:2006za} with default settings.  For signals with a neutralino LSP, we use the MSSM model built into {\tt MadGraph5}.  For signals with a gravitino LSP, we use the publicly available {\tt stopnlsp} UFO model~\cite{stopnlsp,Degrande:2011ua}.  In all simulations, we set $m_t = 172.5$~GeV and $\Gamma_t = 1.3$~GeV.

Signal samples are treated as simple $2\to2$ processes followed by 3- or 4-body decays into $bl^+\nu\neu$ or $ bl^+\nu\gold$, with cross sections normalized to the NLO+NLL predictions for 7~TeV~\cite{Beenakker:2010nq}.  For 8~TeV running, we extrapolate using the leading-order ratios from {\tt MadGraph5}, which agree with NLO ratios derived from {\tt PROSPINO}~\cite{Beenakker:1996ed}.  We set stop widths to 0.01~GeV.  All other SUSY particles are decoupled.  For a small subset of models, we have generated samples matched up to one extra jet in production, and found no significant differences in the relevant kinematic distributions.  This gives some indication that our claimed final signal efficiencies are under good theoretical control, at least on the production side.  It would also be useful to have a more systematic treatment of stop decays beyond leading-order.  We note that some degree of matrix element and shower matching in the decay is automatically provided by {\tt PYTHIA}~\cite{Sjostrand:2006za}, when the top quark appears as an on-shell particle in the {\tt MadGraph5} event record.

For backgrounds other than $t\bar t$, we often match up to one or two extra jets, using the 5-flavor cone-MLM prescription~\cite{Caravaglios:1998yr} with $p_T = 20$~GeV and $\Delta R = 0.3$.  We apply $K$-factors to rescale the cross sections to their NLO values (which can be found, for example, in~\cite{Aad:2012uu}).  The background samples include:
\begin{itemize}
\item  $tW$ matched up to one extra jet ($K = 1.15$).
\item  $t\bar t W$ and $t\bar t Z$ ($K = 1.3$).
\item  $W^+W^-$ matched up to two extra jets ($K = 1.4$).
\item  $WZ$ and $ZZ$ ($K = 1.4$).
\item  $l^+l^-$+jets matched up to two extra jets ($K = 1.3$).
\end{itemize}

After showering and hadronization, we reconstruct events in the manner described in~\cite{Aad:2012uu}, using {\tt FastJet} {\tt v2.4.2} for jet reconstruction~\cite{Cacciari:2005hq}, and smearing final object energies and directions by-hand without an explicit detector model.  Jets are defined with the anti-$k_T$ algorithm with radius $R = 0.4$~\cite{Cacciari:2008gp}.  We smear the jet energies by a relative fraction $(5\;{\rm GeV})/E \oplus (0.5\;{\rm GeV}^{1/2})/\sqrt{E} \oplus 0.04$, and $\eta/\phi$ individually by $0.025$.  We smear electron energies by $0.02$, and muon energies by $(0.1)\times\sqrt{E/{\rm TeV}}$.  We define $\vecMET$ as the 2-vector momentum sum of all final-state invisible particles (neutrinos and LSPs).  We smear the individual $x$ and $y$ components of this vector by ($0.7$~GeV$^{1/2}$)$\times\sqrt{H_T}$, where $H_T$ is the scalar-summed $p_T$ of all visible particles with $|\eta| < 5$.\footnote{Since we include {\it all} particles, the amount of smearing is sensitive to the shower and underlying event models.  This is not a major issue as long as all samples use the same models.  However, the {\tt MC@NLO} $H_T$ spectrum for $t\bar t$ comes out about 10\% harder than the {\tt MadGraph5} $H_T$ spectrum for $2\to 2$ $t\bar t$.  Nonetheless, we have checked that a 10\% rescaling of $H_T$ has only a minor impact on the final smeared $m_{T2}$ distributions.}  We calibrated this coefficient within our own simulation such that the $m_{T2}$ spectrum for $t\bar t$ matches ATLAS's simulation, which in turn agrees well with the data when added to the other backgrounds.  The coefficient also matches ATLAS's most recent measurement of $\MET$ resolution, based on 2011 data~\cite{ATLASMET}.  (Note that the resolution degrades with pileup, but that subtraction methods should actually be capable of stabilizing the coefficient at $0.5$~GeV$^{1/2}$.)  To match the normalization of the ATLAS $t\bar t$ simulation, we reweight all events by an efficiency factor of $0.8$.  This effectively incorporates effects such as inefficiencies in isolated lepton identification.

We identify a jet's flavor by matching in $\eta$-$\phi$ space to bottom- and charm-flavored hadrons in the event record.  The assigned flavor is then the heaviest one found.  We apply $b$-tagging efficiencies of 0.6, 0.1, and 0.01 for $b$-jets, $c$-jets, and unflavored-jets, respectively.  We have found that our relative rates of $b$-tagged and untagged $t\bar t$ events are in good agreement with ATLAS.

\section{Details of Statistical Procedures}
\label{sec:statistics}

Our starting point is the signal and aggregated background cross section predictions in the control region (CR) and signal region (SR):  $\sigma_{\rm S,CR}$, $\sigma_{\rm S,SR}$, $\sigma_{\rm B,CR}$, $\sigma_{\rm B,CR}$.  From these, we define relative efficiencies to fall into the SR,
\begin{equation}
\epsilon_{\rm S} \,\equiv\, \frac{\sigma_{\rm S,SR}}{\sigma_{\rm S,CR}}, \;\;\; \epsilon_{\rm B} \,\equiv\, \frac{\sigma_{\rm B,SR}}{\sigma_{\rm B,CR}}.
\end{equation}
We also define the signal fraction in the CR,
\begin{equation}
f_{\rm CR} \,\equiv\, \frac{\sigma_{\rm S,CR}}{\sigma_{\rm B,CR}}.
\end{equation}
For a given ``observed'' CR count $N_{\rm CR}$, we can then predict what would be expected in the SR for the background-only hypothesis and the signal-plus-background hypothesis:
\begin{equation}
B_{\rm SR} \,=\, \epsilon_{\rm B} N_{\rm CR}, \;\;\; (S+B)_{\rm SR} \,=\, \frac{\epsilon_{\rm B} + \epsilon_{\rm S}f_{\rm CR}}{1+f_{\rm CR}} N_{\rm CR}.
\end{equation}
Our test statistic is the ``observed'' SR count $N_{\rm SR}$.  Note that this method automatically loses discrimination power if $\epsilon_S = \epsilon_B$, in which case a realistic experiment would indeed normalize-away the signal.

We determine the exclusion potential of the signal using the $CL_S$ method~\cite{Read:2002hq}.  Based on $B_{\rm SR}$ and $(S+B)_{\rm SR}$, we run Poissonian pseudo-experiments to build up the discrete probability density functions (p.d.f.'s) in $N_{\rm SR}$.  We fix $N_{\rm CR}$ to its expectation value for background-only for a given luminosity.  (The effects of introducing Poisson fluctuations into $N_{\rm CR}$, or of instead setting it according to the $(S+B)$ expectation, are small and are not considered.)  The fact that the quantiles are discrete leads to a minor ambiguity over which bins to include in the $CL_S$ calculation.  We always err on the conservative side, and pick the weaker of the two choices.

In order to account for systematic errors on the background modeling, we also allow $\epsilon_{\rm B}$ to fluctuate in our pseudoexperiments.  We typically assume 15\% gaussian error.  We do not apply a systematic error on the signal efficiency or $f_{\rm CR}$.  ($\epsilon_S$ is typically much less affected by instrumentation effects than $\epsilon_B$, and those effects are correlated between the two $\epsilon$'s.  $f_{\rm CR}$ is affected by theoretical uncertainties in the total cross section, which we briefly discuss in section~\ref{sec:results2012}.)

As a possible extension of our search, we also study a full shape-based analysis with vanishing systematic errors.  These compare binned $m_{T2}$ distributions (after the $\MET/m_{\rm eff}$ cut) of pure background and signal-plus-background.  While we include all backgrounds in this analysis, we separately rescale the signal added to the top-like background ($t\bar t$, $tW$, and $t\bar t W/Z$) to match the normalization of the top-like background alone.  Our test statistic is the log-likelihood ratio of $B$ and $S+B$ hypotheses over all bins (essentially the sum of the measured bin counts, weighted individually by $(S+B)_i/B_i-1$).  We construct the p.d.f.'s in this test statistic using sets of pseudo-experiments distributed according to either $B$ or $S+B$.

Some of our results claim very high exclusion potential, for example in excess of $5\sigma$.  Estimating $CL_S$ in these cases becomes impractical using a pseudo-experiment approach.  In practice, we use $10^5$ pseudo-experiments to generate each p.d.f.  When our $CL_S$ computation involves integrals that contain less than 5 events, we switch to a gaussian model.  When they contain between 5 and 50 events, we average the exclusions (expressed in $\sigma$'s) from the two procedures.

For setting discovery thresholds, we determine whether the median of the $S+B$ p.d.f.~is separated by more than $5\sigma$ from the median of the $B$ p.d.f., measured in standard deviations of the latter.  We always use the gaussian approximation for this calculation.  Usually, the threshold for $5\sigma$ discovery has a close correspondence with a particular $CL_S$ exclusion level, and we have exploited this correspondence in constructing the discovery line in Fig.~\ref{fig:excl8grav}.

Note that our treatment simply sums over same-flavor and different-flavor channels.  A more comprehensive analysis would treat them independently.  In particular, some model points with off-shell decays may have low efficiency for $b$-jet reconstruction, which would deplete the same-flavor channel's $S/B$ due to our $b$-tag requirement.


\bibliography{lit}
\bibliographystyle{apsper}

\end{document}